\documentclass[12pt,preprint]{aastex}
\slugcomment{\bf 2007 March $9^{th}$}
\shorttitle{The Massive Hosts of Radio Galaxies Across Cosmic Time}
\shortauthors{Seymour et al.}

\def\arcsec{\ifmmode {^{\prime\prime}}\else $^{\prime\prime}$\fi}
\def\arcmin{\ifmmode {^{\prime}}\else $^{\prime}$\fi}
\def\farcs{\ifmmode \rlap.{^{\prime\prime}}\else
    $\rlap.{^{\prime\prime}}$\fi}
\def\arcmper{\ifmmode \rlap.{^{\prime}}\else
    $\rlap.{^{\prime}}$\fi}

\def\lya{\ifmmode {\rm Ly\alpha}\else{\rm Ly$\alpha$}\fi}
\def\Lya{\ifmmode {\rm Ly\alpha}\else{\rm Ly$\alpha$}\fi}
\def\LFIR{\ifmmode {\rm \,L_{FIR}}\else ${\rm \,L_{FIR}}$\fi}
\def\Lsun{\ifmmode {\rm\,L_\odot}\else ${\rm\,L_\odot}$\fi}
\def\Msun{\ifmmode {\rm\,M_\odot} \else ${\rm\,M_\odot}$\fi}
\def\Msunpyr{\ifmmode {\rm\,M_\odot\,yr^{-1}} \else {${\rm\,M_\odot\,yr^{-1}}$}\fi}
\def\pyr{\ifmmode {\rm\,yr^{-1}} \else {${\rm\,yr^{-1}}$}\fi}
\def\kms{\ifmmode {\rm\,km~s^{-1}} \else ${\rm\,km\,s^{-1}}$\fi}
\def\kmps{\ifmmode {\rm\,km~s^{-1}} \else ${\rm\,km\,s^{-1}}$\fi}
 
\def\ergps{\ifmmode {\rm\,erg\,s^{-1}} \else {${\rm\,erg\,s^{-1}}$}\fi}
\def\ergpspcm{\ifmmode {\rm\,erg\,s^{-1}\,cm^{-2}} \else {${\rm\,erg\,s^{-1}\,cm^{-2}}$}\fi}
\def\surfbr{\ifmmode {\rm\,erg\,s^{-1}\,cm^{-2}\,arcsec^{-2}} \else {${\rm\,erg\,s^{-1}\,cm^{-2}\,arcsec^{-2}}$}\fi}
\def\spose#1{\hbox to 0pt{#1\hss}}
\def\simlt{\mathrel{\spose{\lower 3pt\hbox{$\mathchar"218$}}
     \raise 2.0pt\hbox{$\mathchar"13C$}}}
\def\simgt{\mathrel{\spose{\lower 3pt\hbox{$\mathchar"218$}}
     \raise 2.0pt\hbox{$\mathchar"13E$}}}
\def\cf{{\it c.f.,}}
\def\eg{{\it e.g.,}}
\def\ie{{\it i.e.,}}

\def\erg{{\rm\thinspace erg}}

\def\uJy{{\rm\thinspace \mu Jy}}

\def\km{{\rm\thinspace km}}

\def\Mpc{{\rm\thinspace Mpc}}

\def\s{{\rm\thinspace s}}

\def\ergps{\mbox{$\erg\s^{-1}$}}

\def\kmps{\hbox{$\km\s^{-1}\,$}}

\def\kmpspMpc{\hbox{$\km\s^{-1}\Mpc^{-1}\,$}}

\def\um{\hbox{$\mu {\rm m}$}}

\def\Msun{{\rm\thinspace M_\odot}}
\begin{document}

\title{The Massive Hosts of Radio Galaxies Across Cosmic Time}

\author{Nick Seymour\altaffilmark{1},
Daniel Stern\altaffilmark{2},
Carlos De Breuck\altaffilmark{3},
Joel Vernet\altaffilmark{3},
Alessandro Rettura\altaffilmark{4},
Mark Dickinson\altaffilmark{5},
Arjun Dey\altaffilmark{6},
Peter Eisenhardt\altaffilmark{2},
Robert Fosbury\altaffilmark{3},
Mark Lacy\altaffilmark{1},
Pat McCarthy\altaffilmark{6},
George Miley\altaffilmark{7},
Brigitte Rocca-Volmerange\altaffilmark{8},
Huub R\"ottgering\altaffilmark{9},
S. Adam Stanford\altaffilmark{10,11},
Harry Teplitz\altaffilmark{1},
Wil van Breugel\altaffilmark{10,11}
\& Andrew Zirm\altaffilmark{4}}

\altaffiltext{1}{{\it Spitzer} Science Center, California Institute of
  Technology, 1200 East California Boulevard, Pasadena, CA 91125, USA. 
[email: {\tt seymour@ipac.caltech.edu}]}
\altaffiltext{2}{Jet Propulsion Laboratory, California Institute of Technology, 
Pasadena, CA 91109, USA.}
\altaffiltext{3}{European Southern Observatory, Karl Schwarzschild Stra\ss e, 
D-85748 Garching, Germany.}
\altaffiltext{4}{Johns Hopkins University, 3400 N. Charles Street, Baltimore, 
MD 21218, USA.}
\altaffiltext{5}{National Optical Astronomy Observatory, Tucson, AZ 85719, USA.}
\altaffiltext{6}{Carnegie Observatories, 813 Santa Barbara Street, Pasadena, 
CA 91101, USA.}
\altaffiltext{7}{Leiden Observatory, University of Leiden, PO Box 9513, 
2300 RA Leiden, Netherlands.}
\altaffiltext{8}{Institute d'Astrophysique de Paris, 98bis Bd Arago, 75014 Paris, 
France.}
\altaffiltext{9}{University of California, Davis, CA 95616, USA.}
\altaffiltext{10}{Institute of Geophysics and Planetary Physics, 
Lawrence Livermore National Laboratory, Livermore, CA 94551, USA.}
\altaffiltext{11}{University of California, Merced, PO Box 2039, Merced, 
CA 95344, USA.}

\begin{abstract}

We present the results of a comprehensive {\it Spitzer} survey of 69
radio galaxies across $1 < z < 5.2$.  Using IRAC ($3.6 - 8.0\, \mu$m),
IRS ($16\, \mu$m) and MIPS ($24 - 160\, \mu$m) imaging, we decompose the
rest-frame optical to infrared spectral energy distributions into stellar,
AGN, and dust components and determine the contribution of host galaxy
stellar emission at rest-frame $H-$\,band.  Stellar masses derived from
rest-frame near-IR data, where AGN and young star contributions are 
minimized, are significantly more reliable than those derived from 
rest-frame optical and UV data.  We find that the fraction of emitted 
light at rest-frame $H-$\,band from stars is $>60\%$ for $\sim75\%$ the 
high redshift radio galaxies.
As expected from unified models of AGN, the stellar fraction of the
rest-frame $H-$\,band luminosity has no correlation with redshift, radio
luminosity, or rest-frame mid-IR ($5\,\mu$m) luminosity. Additionally,
while the {\em stellar} $H-$\,band luminosity does not vary with stellar
fraction, the {\em total} $H-$\,band luminosity anti-correlates with
the stellar fraction as would be expected if the underlying hosts of
these radio galaxies comprise a homogeneous population. The resultant
stellar luminosities imply stellar masses of $10^{11-11.5}\, M_\odot$ 
even at the highest redshifts. Powerful radio
galaxies tend to lie in a similar region of mid-IR color-color space
as unobscured AGN, despite the stellar contribution to their mid-IR SEDs at 
shorter-wavelengths. The mid-IR luminosities alone classify most 
HzRGs as LIRGs or ULIRGs with even higher total-IR luminosities. As 
expected, these exceptionally high mid-IR luminosities
are consistent with an obscured, highly-accreting AGN. We find a weak
correlation of stellar mass with radio luminosity.

\end{abstract}


\section{Introduction}

Luminous radio galaxies were the first class of obscured, or {\it type
2}, quasars to be discovered and characterized. They have accreting
super-massive black holes whose continuum emission at UV, optical,
and soft X-ray energies is absorbed by dust, thus allowing a {\em clear} 
view of the host galaxy. More
recently, hard X-ray and mid-IR surveys have identified the radio-quiet
cousins to luminous radio galaxies \citep[\eg][]{Norman:02,Stern:02a,
MartinezSansigre:05, Polletta:06, Lacy:06}.  The main evidence
that radio galaxies host super-massive black holes comes from the
high luminosities of their radio lobes, which are fed by radio jets
originating at the host galactic nuclei.  The lobe spatial extents
\citep[up to a few Mpc, ][]{Saripalli:05} and luminosities ($L_{1.4{\rm
GHz}} \ge 10^{25}$\,W\,Hz$^{-1}$) clearly rule out a stellar origin for
their energetics \citep[\eg][]{Rees:78}.  In terms of the orientation
unification scheme for AGN \citep[\eg][]{Barthel:89,Antonucci:93,Urry:95},
radio galaxies are radio-loud quasars seen from an angle where an
optically-thick torus obscures emission from the region closest to the
central engine.

\begin{figure}
\includegraphics[width=13cm,angle=0]{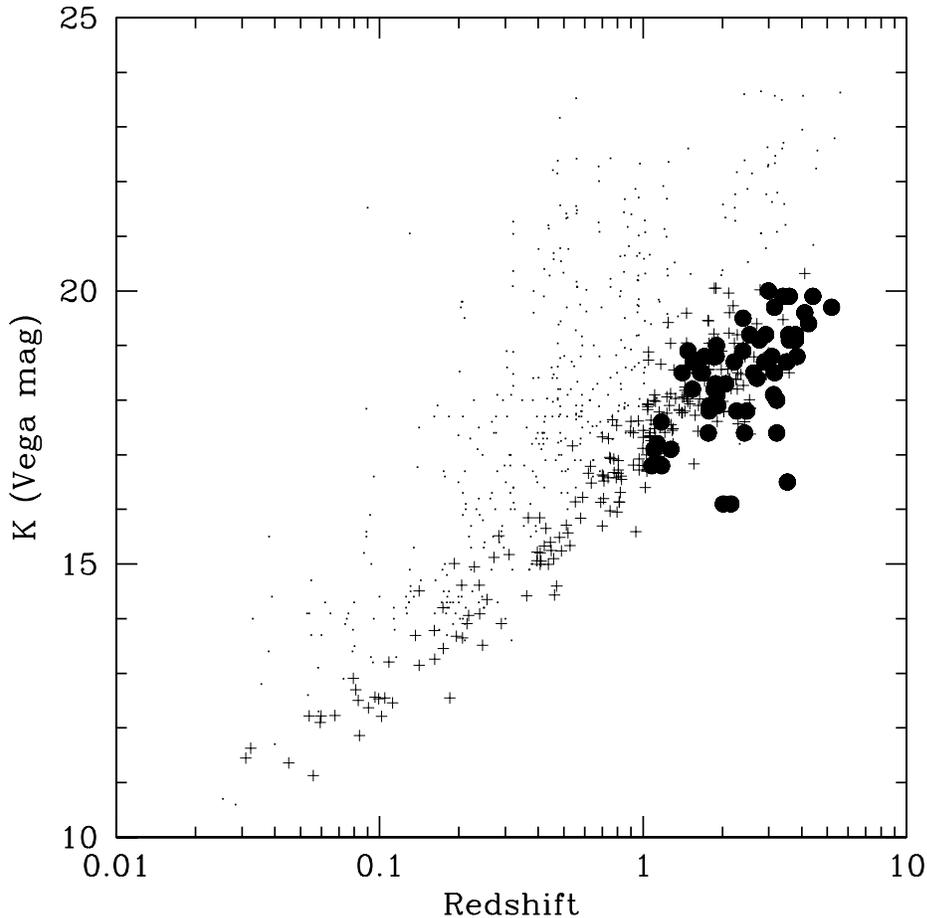}
\caption{Hubble $K-z$ diagram of radio-loud (pluses and filled circles)
and radio-quiet galaxies (points).  The radio-loud sample is from
De Breuck et al. (2002).  The radio quiet samples are from the Hawaii
survey (Songaila et al. 1994) and the {\it Hubble} Deep Field - North
(Dickinson et al. 2003).  The filled circles indicate the 69 $z > 1$
radio galaxies which comprise our {\it Spitzer} sample.  The three HzRGs
falling below the obvious correlation are MRC 0156-252 ($z=2.016$), 
PKS 1138-262 ($z=2.156$)and TX J1908+7220 ($z=3.530$).
MRC 0156-252 is known to have strong H$\alpha$ in the $K$-band 
(Eales \& Rawlings 1996) and PKS 1138-262 has also recently been confirmed 
to have strong H$\alpha$ in the $K$-band (Nesvadba et al. 2006). TX J1908+7220
has a strong unresolved AGN component (see notes in individual sources)
and seems closer to being an unobscured quasar than a radio galaxy.
\label{fig.Kz}}
\end{figure}

Due to their large radio luminosities, radio galaxies were the predominant
way to probe the distant universe until the advent of $8 - 10$\,m class
telescopes and the Lyman-break technique in the last decade.  In fact,
radio galaxies were the first galaxies to be found above redshifts one,
two, three and four \citep[see][and references therein]{Stern:99e}.
Across cosmic time, the host galaxies of powerful radio sources appear
to be uniquely robust indicators of the most massive galaxies in the
universe.  At low redshift this result has been known since the first
optical identifications of extra-galactic radio sources showed them
to be associated with massive, giant elliptical (gE and cD) galaxies
\citep{Matthews:64}. In the more distant universe, indirect evidence that
this association remains intact comes from the detection of host galaxies
with $r^{1/4}$ law light profiles in {\it Hubble Space Telescope} ({\it
HST}) observations of high-redshift radio galaxies (HzRGs) at $1\simlt
z \simlt 2$ \citep{Pentericci:01, Waddington:02, Zirm:03, Bunker:07};
the tendency for HzRGs to reside in moderately rich (proto-)cluster
environments \citep{LeFevre:96, Pascarelle:96, Venemans:02, Venemans:03,
Venemans:04, Venemans:05, Venemans:06, Stern:03}; the spectacular
($>100$\, kpc) luminous \lya\, haloes seen around several sources,
implying large gas reservoirs \citep{Reuland:03, VillarMartin:03}; sub-mm
detections of HzRGs, implying violent star formation activity up to $\sim
1000\, M _{\odot}\, {\rm yr}^{-1}$ \citep{Archibald:01, Reuland:04};
and a few direct kinematic measurements of HzRGs \citep{Dey:96,
Dey:96b,Nesvadba:06}.  The most compelling evidence of this association
of HzRGs with the most massive systems, however, is the tight correlation
of the observed near-infrared Hubble, or $K-z$, diagram for powerful
radio sources \citep[Fig.~\ref{fig.Kz};][]{Lilly:84, Best:98, Eales:97,
vanBreugel:98, Jarvis:01, DeBreuck:02, Willott:03a, RoccaVolmerange:04,
Brookes:06}: HzRGs form a narrow redshift sequence which traces the
envelope of radio-quiet galaxies and is well modeled by the evolution
of a stellar population formed at high redshift from an accumulation of
up to $10^{12}\,M_{\odot}$ of pre-galactic baryonic material. The large-scale,
double-lobed radio morphologies and enormous radio luminosities suggested
early on that HzRGs must have spinning super-massive black holes powering
relativistic jets in their centers \citep{Rees:78, Blandford:82}.
With the more recent discovery that the stellar bulge and central black
hole masses of galaxies are closely correlated \citep[\eg][]{Tremaine:02},
it is no longer a surprise that the parent galaxies of the most powerful
radio sources occupy the upper end of the galaxy mass function.

Despite two decades of study since the initial discovery of the HzRG $K-z$
relation \citep{Lilly:84}, the nature and tightness of the relation
remains mysterious.  The scatter in the relation is surprisingly low
out to $z \sim 1$, and only increases modestly at higher redshifts as
the observed $K$-band samples rest-frame optical emission \citep[e.g.,
Fig.~\ref{fig.Kz};][]{DeBreuck:02}.  Interpretations have generally relied
on these observed near-IR observations probing the stellar populations
of the distant radio galaxies \citep[\eg][]{RoccaVolmerange:04}.
However, detailed interpretations have been complicated by (i) uncertain
contributions of AGN-related light which are most important in the 
rest-frame UV \citep[\eg][]{Vernet:01}, (ii) evolution of the host galaxy
stellar population(s), and (iii) band-shifting, meaning that observed
$2.2\, \mu$m samples very different emitting wavelengths for different
redshift sources, reaching into the rest-frame ultraviolet for the most
distant HzRGs.

In this paper, we present observations of a large sample of HzRGs
obtained with the {\it Spitzer Space Telescope} \citep{Werner:04}.
By observing the same {\it rest-frame} near- to mid-IR spectral range
for sources over a large redshift range ($1 < z < 5.2$), we remove many
of the complications which plagued previous studies.  In particular,
the $1.6\, \mu$m peak of the stellar emission provides a reasonably
robust measure of the stellar mass for stellar populations with ages
$\simgt 1$\,Gyr \citep{Sawicki:01}.  While UV and optical emission has
strong contributions from the youngest and hottest stars in a galaxy,
near-IR emission primarily derives from the low-mass stars which dominate
the stellar mass of a galaxy.  Thus, while rest-frame UV and optical
studies of galaxies are well-suited to probe galaxy star-formation rates,
rest-frame near-IR studies are well-suited to probe galaxy stellar masses.
We discuss this approach and other recent results that might effect it  
\citep[e.g.][]{Maraston:05} in detail in \S6.

\begin{figure}
%\clearpage
\includegraphics[width=13cm,angle=270]{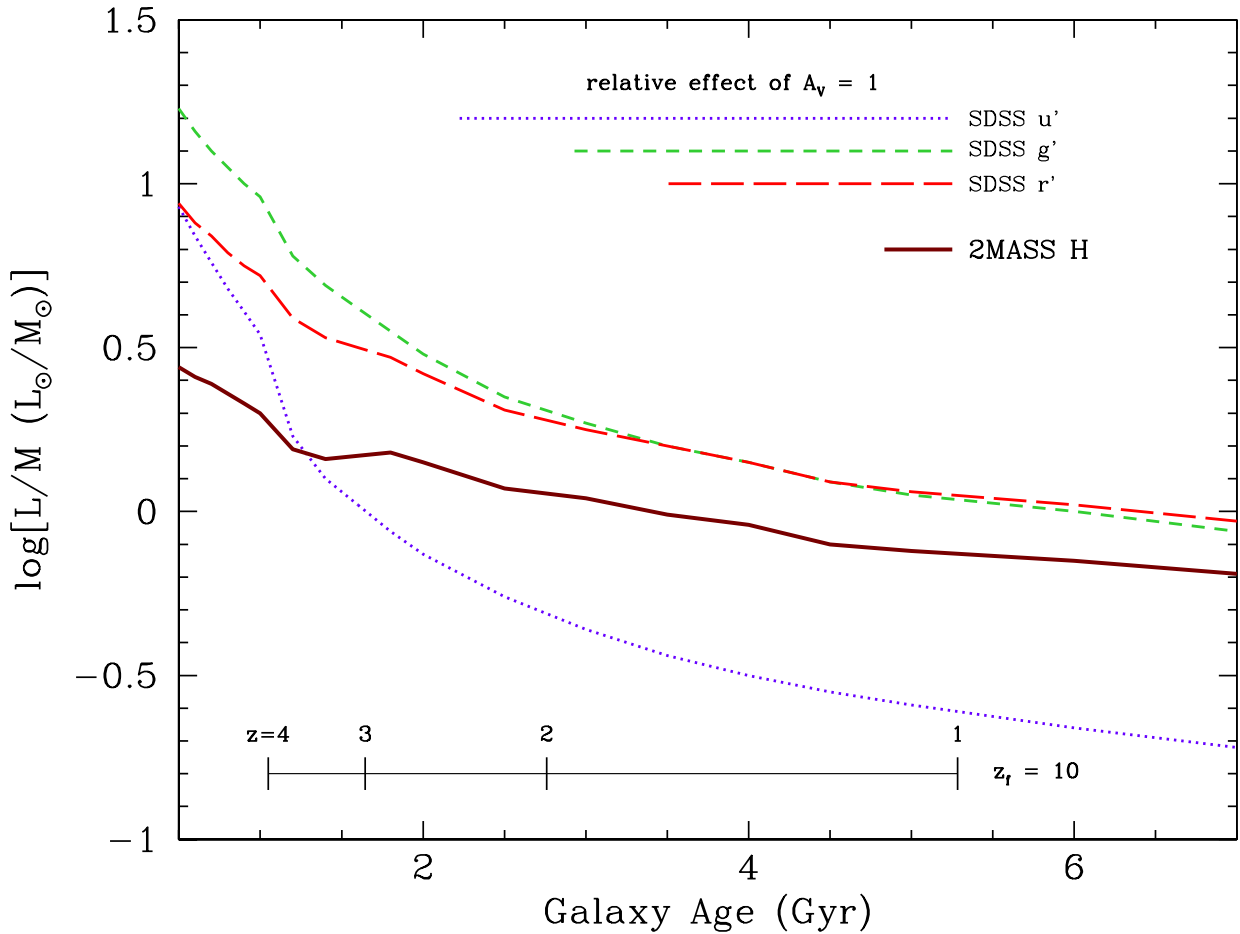}
\caption{The light-to-mass ratios (in solar units) as a function 
of age for the adopted {\tt P\'EGASE.2} elliptical galaxy model for 
different rest-frame optical (SDSS $u'$, $g'$, $r'$) and near-IR (2MASS 
$H$) passbands. The light-to-mass ratio for the $H-$\,band filter, near the 
peak of stellar emission, varies much less as a function of age than 
the shorter wavelength filters.  Additionally, the optical passbands are 
more susceptible to (i) dust absorption --- the relative effects of one 
magnitude of visual extinction in each of the passbands is illustrated by 
the horizontal lines; in units of $A_\lambda$, $H-$\,band is more than four 
times less sensitive to dust extinction than $r'$, (ii) 
uncertainty in the age of the stellar population, and (iii) the addition 
of light from a second, younger population.  For a formation redshift of 
$z_{\rm form}= 10$, the corresponding ages are marked near the bottom of 
the figure for different redshifts.  For $z_{\rm form}=14$ or 7, the 
tracks shift 0.2\,Gyrs to the left or 0.3\,Gyrs to the right,
respectively.  As the $H-$\,band track is fairly flat, this uncertain
formation redshift has little effect on our derived light-to-mass ratio.
\label{fig.m2l}}
\end{figure}

Fig.~\ref{fig.m2l} illustrates the advantage of rest-frame near-IR
derivations of stellar masses.  First, the impact of dust extinction falls 
sharply with wavelength, making quantities derived from rest-frame near-IR
observations more than an order of magnitude less susceptible to uncertain 
dust extinction corrections relative to optical observations.  Second,
since the main sequence lifetimes of low-mass stars exceed the Hubble
time, galaxy masses derived from rest-frame near-IR observations are
relatively insensitive to the age of stellar populations or the star
formation history.  In particular, secondary bursts of star formation
will affect UV and optical magnitudes significantly more than near-IR
magnitudes. 

We here use mid-IR observations of a large sample of radio galaxies
to probe their host galaxy stellar masses.  By observing consistently
on both sides of the $1.6\,\um$ peak of the stellar emission, we
avoid complicated $k-$correction effects and derive a reasonably
robust measure of the stellar mass for stellar populations with
ages $\simgt 1$\,Gyr. Observations at longer wavelengths allow us
to determine the contribution of warm, AGN-heated dust emission to
the rest-frame near-IR emission.  Our paper is organized as follows.
Section~2 describes the {\it Spitzer} HzRG sample.
Section~3 presents the {\it Spitzer} mid- to far-IR data and their
reduction.  Section~4 describes our $\chi^2$ fitting of the spectral
energy distributions (SEDs) and \S5 presents the results of this
fitting. Section 6 describes how the derived rest-frame $H-$\,band
luminosities are converted into stellar masses.  Section~7 discusses
and concludes this analysis.  We present notes on individual sources as
an Appendix.  This paper presents our entire {\it Spitzer} data set and
analyzes the bulk properties of the sample.  {\it Spitzer} observations of
individual sources from this program have been the subject of detailed
studies by \citet[][MRC~2104$-$242 at $z = 2.49$]{VillarMartin:06},
\citet[][LBDS~53W091 at $z = 1.55$]{Stern:06}, \citet[][PKS at 
$z = 2.156$]{Broderick:07} and \citet[][4C~23.56 at $z = 2.48$]{DeBreuck:07}.
Throughout we assume a concordance model\footnote{These common values
of the ``concordance'' cosmology underestimate luminosities (and hence
stellar masses) by $\sim 2 - 3\%$ at $z = 3 - 4$ when compared to
the latest {\it Wilkinson Microwave Anisotropy Probe} measurements,
$\Omega_M = 1 - \Omega_{\Lambda} = 0.25$, $\Omega_0 = 1$, and $H_0 =
73\, \kmpspMpc$ \citep{Spergel:06}.  Lower-redshift HzRGs show an even
smaller systematic adjustment, implying that the assumed cosmological
parameters are inconsequential to our analysis.} of universe expansion,
$\Omega_M = 1 - \Omega_{\Lambda} = 0.3$, $\Omega_0 = 1$, and $H_0 =
70\, \kmpspMpc$.  Inferred luminosities presented in this paper are of
the form $\nu L_\nu/L_\odot$, where $L_\odot=3.9\times10^{23}$\,W.

\section{The {\it Spitzer} High-Redshift Radio Galaxy Sample}

Our HzRG sample is drawn from radio galaxy surveys executed during the
last 45 years \citep[starting with the 3CR;][]{Bennett:61}.  We have
searched both flux-limited surveys such as the 3C \citep{Spinrad:85}, 6CE
\citep{Eales:97}, 7C \citep{Lacy:99,Willott:01}, MG \citep{Bennett:86,
Lawrence:86}, and MRC \citep{McCarthy:96}, as well as surveys filtered
by their ultra-steep radio spectra \citep[USS; \eg][]{Chambers:96,
Rottgering:97, DeBreuck:00}.  The former provide a sample unbiased in
their radio properties, but lack the high redshift ($z > 2$) sources
mainly found in the USS samples.  Because the USS samples also probe
fainter radio flux densities, including such sources allows us to break
the strong luminosity$-$redshift degeneracy of flux density-limited
surveys.

For the purposes of this work we define a HzRG as a radio galaxy above
a redshift of one with a rest-frame 3\,GHz luminosity greater than
$10^{26}$\,WHz$^{-1}$. This choice is to ensure that we include only
those objects with very powerful obscured AGNs. This value
is similar to that used by other authors to separate AGN based upon
their radio luminosities.  The classical luminosity that separates
\citet{Fanaroff:74} type~1 and type~2 sources (which are morphologically
distinct), converted from 178\,MHz to 3\,GHz, is $10^{25.4}$
($10^{24.8}$)\,W\,Hz$^{-1}$ assuming a steep (ultra-steep) radio spectral 
index of $\alpha=-0.8$ ($-1.3$).  A more recent radio-loud/radio-quiet 
division is $L_{\rm 3GHz}=10^{26.3}$ ($10^{26.4}$)\,W\,Hz$^{-1}$ 
\citep{Miller:90} and an upper limit to radio-quiet QSOs of 
$L_{\rm 3GHz}=10^{25.2}$ ($10^{25.1}$)\,W\,Hz$^{-1}$ was found by 
\citet{Gregg:96}, both converted to 3\,GHz luminosity assuming the same 
spectral index values.

\begin{figure}
\includegraphics[width=13cm,angle=0]{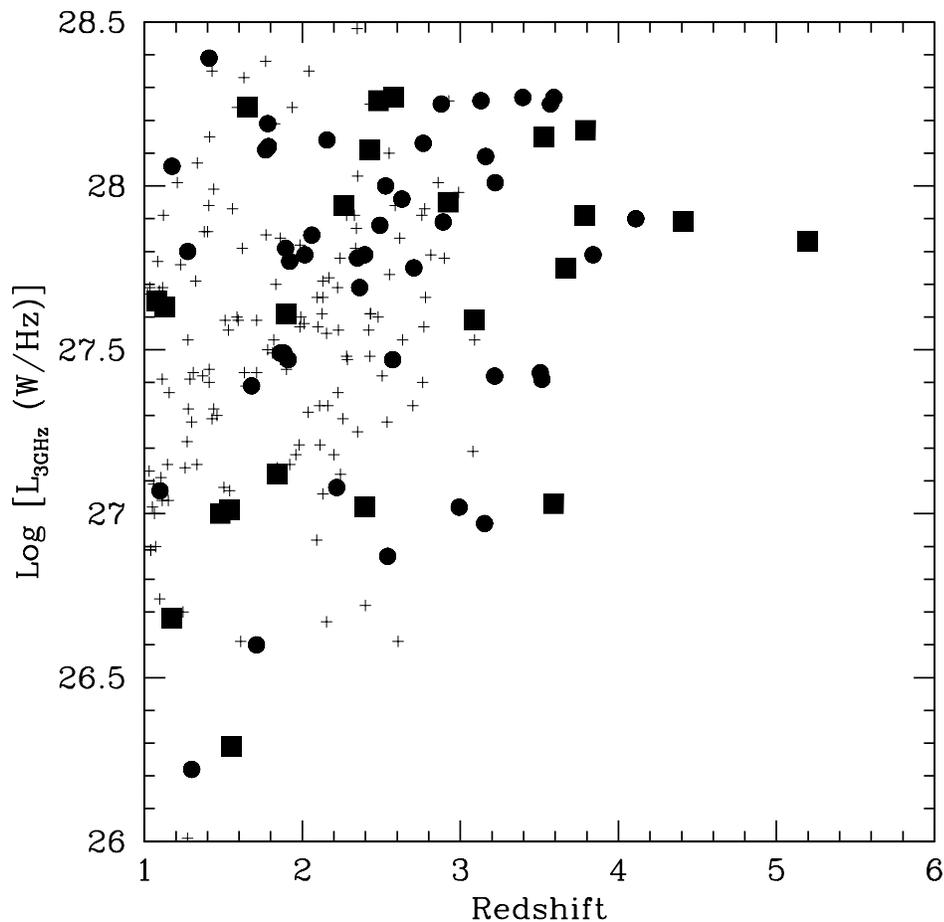}
\caption{Radio luminosity at rest-frame 3\,GHz plotted against redshift for
263 $z > 1$ radio galaxies from the literature (pluses).  Filled symbols
indicate sources in our {\it Spitzer} program, carefully chosen to
uniformly cover $1.5-2$ orders of magnitude in radio power at each redshift
$1 < z < 4$.  The entire {\it Spitzer} sample was observed with IRAC.
Radio galaxies at $z > 2$ were imaged at $16\, \mu$m with IRS.
Radio galaxies in ``low'' Galactic infrared background (see \S3) are
indicated as solid squares and were observed with MIPS. 
\label{fig.Lz}}
\end{figure}

To examine how {\it Spitzer}-derived quantities depend on redshift and radio 
luminosity, we selected a representative subset of 69 targets from the parent 
sample of 263 known $z > 1$ HzRGs (circa 2003; Fig.~\ref{fig.Lz}).  
Table~\ref{table.shzrg} lists the selected subset, which spans the full 
range of redshifts from $z = 1$ to the redshift of the most distant known 
radio galaxy, TN~J0924$-$2201 at $z = 5.19$ \citep{vanBreugel:99a}.  
Most sources were observed as part of {\it Spitzer} Cycle~1 (Program ID 
number = 3329); a few observations derive from Guaranteed Time Observations 
by {\it Spitzer} instrument teams. We required at least 15 sources in each 
$\Delta z = 1$ bin, covering a range of rest-frame 3\,GHz radio 
luminosity, $L_{\rm 3GHz}$. To obtain a uniform set of radio luminosities of 
a sample of radio galaxies covering the entire sky, we obtained radio flux 
densities from the largest all-sky radio surveys, i.e., the 1.4 GHz NRAO VLA 
Sky Survey \citep[NVSS:][]{Condon:98}, the 325/352 MHz Westerbork Northern
Sky Survey \citep[WENSS:][]{Rengelink:97} and the 365 MHz Texas Survey
\citep{Douglas:96}.  These surveys all have $\sim 1$ arcmin spatial
resolution, which has the advantage that the entire sample appear
essentially as point sources, thereby minimizing issues of missing flux
from over-resolved emission.  We chose $3\,$GHz as the fiducial rest
frequency as it is the mid-point between the survey frequencies at
the median redshift of our sample. This choice ensures that almost all radio
luminosities are interpolated, rather than extrapolated as is often
required when deriving lower rest frequency luminosities (\eg\, 151
or 178\,MHz). \citet{Blundell:98} argue, however, that low frequency
luminosities are a better measure of the total radio luminosity,
as the emission at low frequency is dominated by the steep spectrum,
isotropic radio lobes, while high-frequency emission may be contaminated
by Doppler-boosted emission from flat-spectrum radio cores. Until the
completion of all-sky $<$100 MHz surveys, such low frequency rest-frame 
luminosities cannot be uniformly derived for our sample. 
However, most HzRGs in our sample have weak radio cores, which 
although flatter than the ultra-steep spectrum radio lobes, 
still have steep radio spectra \citep[\eg][]{Athreya:97}, and thus are 
not expected to significantly boost the rest-frame 3 GHz luminosity.

We have preferentially selected sources with the most supporting data.
Forty-seven of the 69 HzRGs in our sample have {\it HST} data,
representing over 1.5\,Ms of {\it HST} time.  We also preferentially
included sources previously selected for guaranteed time observations
(GTO) by the {\it Spitzer} instrument teams, thereby minimizing our
{\it Spitzer} observatory request while maintaining a uniformly selected
sample.  Finally, we preferentially selected sources with millimeter and 
sub-millimeter
observations in order to have information on the SEDs longward of the
{\it Spitzer} coverage.  Note that the sample does not suffer a bias
toward cold dust properties as we select only on the basis of (sub-)mm
observations, not detections.

\section{Infrared Observations and Reductions}

The {\it Spitzer} observations consist of imaging photometry in eight
bands, using the entire instrument complement of the observatory.
Data from the Infrared Array Camera \citep[IRAC;][]{Fazio:04a} comes from
all four bands (3.6, 4.5, 5.8 and $8\, \mu$m --- synonymously referred
to as channels 1 to 4).  The entire sample of 69 HzRGs was
observed with IRAC\footnote{Although 3C65 was observed in only two bands
due to an astrometric error in NED which we have since had corrected.}.
Data from the InfraRed Spectrograph \citep[IRS;][]{Houck:04} obtained in
``peak-up imaging'' mode \citep[\cf][]{Teplitz:05} provides flux densities
at $16\, \mu$m for the 46 objects in our sample at $z > 2$.  For the
lower redshift portion of our sample, IRAC $8\, \mu$m images adequately
cover the SED longward of the $1.6\, \um$ stellar bump.  Finally, the
Multi-band Imaging Photometer for {\it Spitzer} \citep[MIPS;][]{Rieke:04}
was used to image a subset of our sample using all three available bands
(24, 70 and $160\, \mu$m).  Due to the uncertain ability of MIPS to
image against the Galactic infrared background at the time of our GO
Cycle 1 submission, only the 26 HzRGs from our sample lying in ``low''
background regions ($S_{24 \mu{\rm m}} < 20\, {\rm MJy\,sr^{-1}}$)
were observed with MIPS.  Table~\ref{table.shzrg} presents the 69
HzRGs comprising our sample, along with their coordinates, redshifts,
rest-frame 3\,GHz radio luminosities, 64\,kpc $K$-band magnitudes (Vega)
and which {\it Spitzer} instruments they have been observed with.

The 141 unique target/instrument observations comprising our survey
were released from the {\it Spitzer} archive between October 2004 and
June 2006.  Fourteen observations were part of the initial GTO programs
\citep[\eg][]{Stern:06}. In general, our data reduction begins with
the basic calibrated data (BCD) output of the {\it Spitzer} Science
Center pipeline, using the most current version number at the time
the data were initially released (\eg\, S9 to S13). Improvements over
the generations of pipelines implemented here were minor and did not
warrant reprocessing data.  The most useful augmentation was the addition
of a global world coordinate system to the IRS peak-up images in S11.
Note that the units of {\it Spitzer} BCD images are in surface brightness
units of MJy ster$^{-1}$, where 1 MJy ster$^{-1}$ corresponds to $23.5
\uJy$ per square arcsecond.

\subsection{IRAC Observations}

Most of the IRAC observations consisted of four dithered 30\,s exposures in
each of the four IRAC channels.  For each channel, the BCDs were mosaiced
using the {\tt MOPEX} package \citep{Makovoz:05} from the {\it Spitzer}
Science Center and re-sampled by a factor of two to give a pixel scale
of 0\farcs61.  The {\tt MOPEX} outlier (\eg\ cosmic ray, bad pixel)
rejection was optimized for the regions of deepest coverage in the
center of the maps, corresponding to the location of the targeted HzRGs.
The eight GTO IRAC observations were all deeper than our standardized
program and were reduced in a similar manner with {\tt MOPEX}. Source
extraction was performed with {\tt SExtractor} \citep{Bertin:96} in dual
image mode, using the $3.6\, \mu$m channel for source detection. All
IRAC BCD images have astrometry derived from 2MASS survey and are good
to $<0\farcs5$ so we can be confident of the radio galaxy identifications.

We generally used the a 7\arcsec\ diameter aperture for photometry, though
in a few cases where the source was faint and/or had a close companion,
we extracted photometry using a 3\farcs5 diameter aperture.  The aperture
corrections, applied after the conversion from surface brightness to flux
density, are listed in Table~\ref{table.irsmips} and are taken from the
{\it Spitzer} First Look Survey \citep{Lacy:05}.  The resulting flux
densities are listed in Table~\ref{table.spitzer} and include the total
uncertainty in the IRAC photometry; \ie\ the statistical uncertainties
are added in quadrature to the systematic uncertainty of $\sim10\%$. The 
systematic uncertainty is the total uncertainty from several different 
effects: the uncertainty in the IRAC flux density calibration, a flat 
fielding color dependent uncertainty and variation of the PSF across 
the field of view. Non-detections have
$3\sigma$ upper limits listed.  The statistical uncertainties in the mosaiced 
images were calculated from the pixel-to-pixel rms from single BCDs (with 
the native pixel scale), converted to $\uJy$, multiplied by the square root
of the area of the aperture in square pixels, divided by the square root
of the number of frames, and multiplied by the appropriate aperture
correction. In all cases, the radio galaxies are well detected in the
3.6 and $4.5\, \mu$m channels\footnote{3C65 is not observed in the $4.5\,\um$
channel. 6C0140$+$326 is probably detected at $3.6$ and $4.5\,\um$, 
but is blended by a foreground galaxy. Likewise 3C294 is detected in all four 
channels, but the proximity of a bright star means only upper limits can be 
determined for channels at 3.6 and $4.5\,\um$}.  In most cases, the radio galaxies are
also detected in the 5.8 and $8\, \mu$m channels. In no cases were the 
HzRGs resolved in channel 1, the highest resolution of all the {\it Spitzer}
imaging bands with a PSF FWHM of $1\farcs44$.

\subsection{IRS Observations}

The IRS observations consisted of just two nodded exposures of the
$16\,\um$ peak-up imager, each of 61\,s duration. The observations
were executed in {\it Spitzer} Cycle 1, before IRS ``peak-up imaging''
was offered as a supported observing mode. However all the data were
reduced with the more recent S13.2.0 version of the archive pipeline,
which provides astrometry to the BCD products accurate to 0\farcs5
relative to the 2MASS reference frame.  Sub-images were made of just the
blue ($16\,\um$) ``peak-up'' field of view in which the radio galaxy
was centered. Difference images were then created by subtracting one
frame from the other, thereby removing background emission and latent
patterns in the array.

The dithered difference images were then combined with bad pixels
and flaring pixels masked.  Flux densities were extracted with
the {\tt IRAF} {\tt DAOPHOT} task using the parameters listed in
Table~\ref{table.irsmips}. Random uncertainties were determined from
the rms in each BCD and converted to $\uJy$. The total uncertainties in this
table include the random and the $7.8\%$ ($6\%$ zero point uncertainty 
and $5\%$ color correction) systematic uncertainties. The systematic 
and random uncertainties are combined in quadrature and used for the
SED fitting discussed in \S4. These flux densities and uncertainties are 
listed in Table~\ref{table.spitzer}. Non-detections are represented by 
$3 \sigma$ upper limits. All 46 of the {\it Spitzer} HzRG sample at
$z \geq 2$ were observed with IRS, as was MRC~0211$-$256 ($z = 1.300$) due
to an error in the submitted AORs\footnote{It was inadvertently confused
with USS~0211$-$122 ($z = 2.336$).}. We have robust $16\,\mu$m detections
of the radio galaxy in 35 of the 47 observations\footnote{Except for the
non-detection of the highest redshift radio galaxy, TN~J0924$-$2201 ($z =
5.195$), the $16\,\um$ non-detections are evenly distributed across $z
\sim 2 - 4$.}.

\subsection{MIPS Observations}

The MIPS observations were restricted to the 26 HzRGs with
``low'' mid-IR background ($S_{24 \mu{\rm m}} < 20\, {\rm MJy\,sr^{-1}}$)
and consisted of one cycle of 30\,s exposures at $24\, \mu$m, three cycles
of 10\,s exposures at $70\, \mu$m, and five cycles of 10\,s exposures at
$160\, \mu$m.  The $24\, \mu$m BCDs were median filtered to remove a
regular stripe artifact pattern prior to mosaicing.  Both the 24 and $70\,
\mu$m data were then mosaiced with {\tt MOPEX} using standard inputs and
with the pixels re-sampled by a factor of two.  Twenty-two of the $24\,\um$ 
observations resulted in detections, but only four radio galaxies
are detected at $70\, \mu$m.  None of our HzRG sample are detected at
$160\, \mu$m.  As we shall see below, the upper limits at long wavelengths 
are still quite useful for constraining the mid-IR SED. Flux densities were 
extracted  with the {\tt IRAF} {\tt DAOPHOT} task using the parameters listed 
in Table~\ref{table.irsmips}. For the $70\,\um$ band we use an aperture 
correction based on a steep spectrum ($F_\lambda\propto\lambda^{0}$). Again, 
Table~\ref{table.spitzer} provides flux densities and upper limits 
($3\sigma$) for all three MIPS cameras.

The random uncertainties (and $3\sigma$ upper limits in
Table~\ref{table.spitzer}) are derived from the BCDs and converted to
$\uJy$. The color correction can be quite significant for these broad
MIPS bands which were calibrated using Galactic stars, \eg\, typically
for a Raleigh-Jeans spectrum, $F_\lambda\propto\lambda^{-4}$.  For the
70 and 160$\,\um$ bands we include a $5\%$ systematic uncertainty to
account for the uncertain color correction. For 24$\,\um$ we typically
find that the SED is very non-stellar. As a first approximation, we apply
the color correction by multiplying by 0.961 (from the MIPS data 
handbook\footnote{\tt 
http://ssc.spitzer.caltech.edu/mips/dh/mipsdatahandbook3.2.pdf}) for
$\lambda F_\lambda =$ constant, which approximates the observed SEDs
of our HzRGs. We further include a $2\%$ systematic uncertainty for the
remaining unknown color correction. The final systematic uncertainties are
are $4.5\%$, $8.6\%$ and $13\%$ for 24, 70 and 160$\,\um$, respectively, 
including flux calibration uncertainties from the {\it Spitzer} Observers 
Manual and color correction uncertainties described above. 
The systematic uncertainties are combined in quadrature with the statistical 
uncertainties and are presented in Table~\ref{table.spitzer}. 

\subsection{Mid-IR Colors of HzRGs}

Recent work has shown that IRAC mid-IR colors reliably
isolate spectroscopically-confirmed, luminous, unobscured AGN
\citep[\eg][]{Lacy:04, Stern:05b}.  However, since the surface density of
sources with mid-IR colors similar to unobscured AGN is much higher than
that of optically-selected (type 1) quasars, it has been inferred that
obscured, type 2 quasars are likewise identifiable from their mid-IR
colors.  Due to their optical faintness, type 2 quasars are not as well
studied as their bright, type 1 cousins.  However, they are a natural
consequence of AGN unification schemes \citep[\eg][]{Antonucci:93},
and models of the hard X-ray background imply that they outnumber type
1 quasars by factors of several \citep[\eg][]{Treister:04}.

\begin{figure}
\begin{minipage}[l]{8.5cm}
\includegraphics[width=8cm,angle=0]{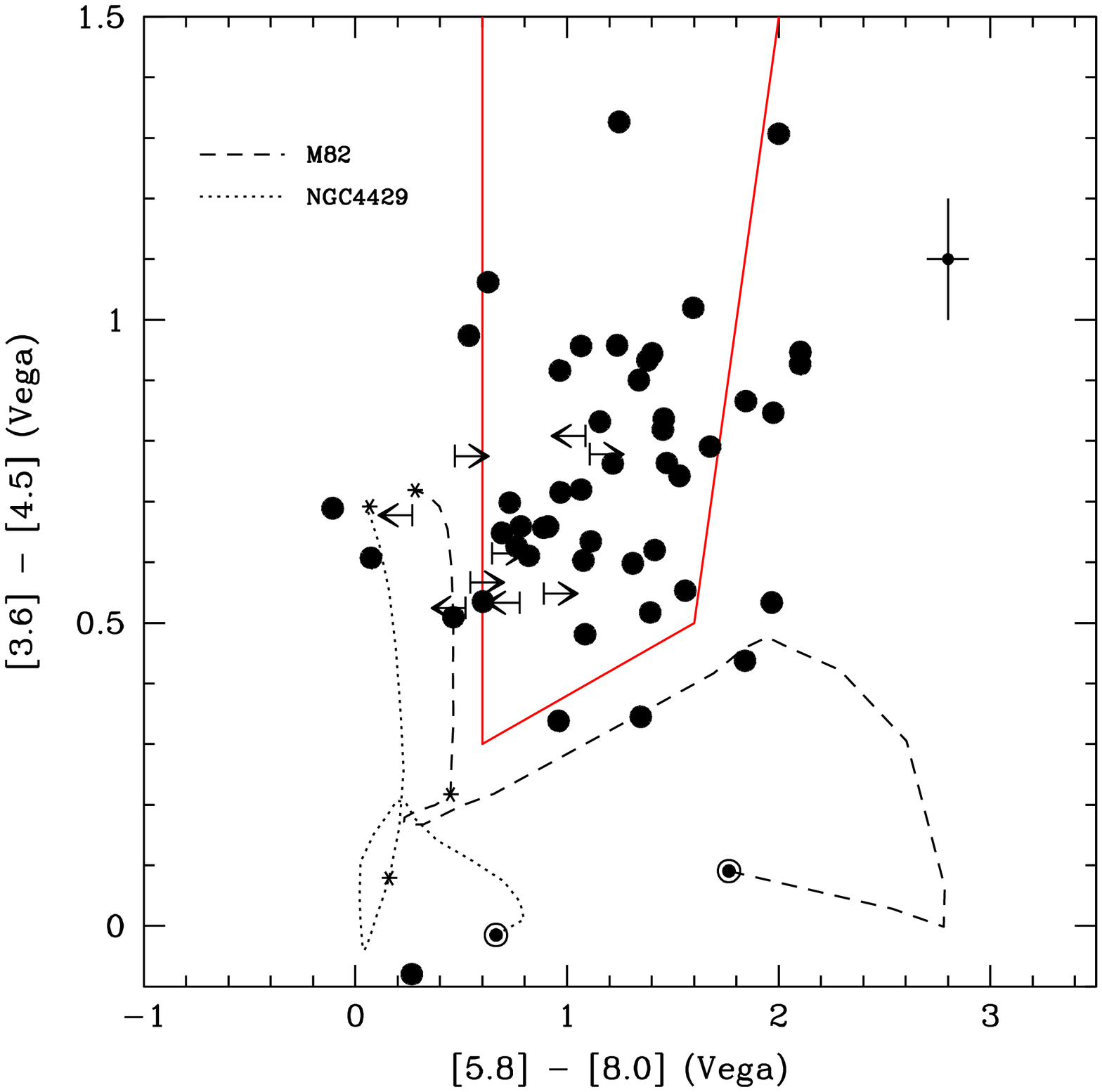}
\end{minipage}
\begin{minipage}[r]{8.5cm}
\includegraphics[width=8cm,angle=0]{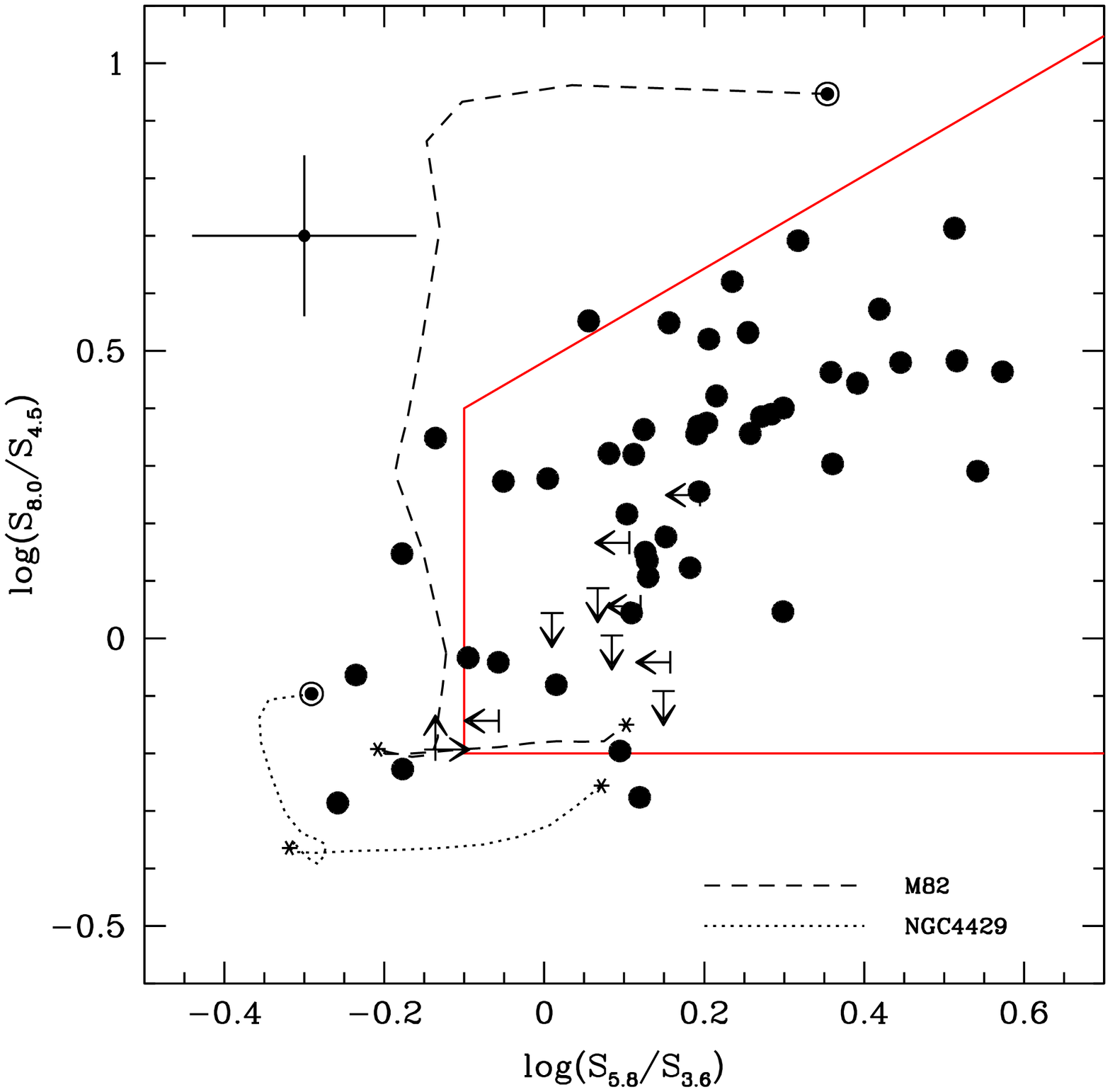}
\end{minipage}
\caption{Mid-IR colors of HzRGs.  Solid lines indicate empirical criteria 
  that separate (type 1) active galaxies from Galactic stars and normal 
  galaxies. The left-hand panel shows the criteria from Stern et al. (2005);
  the right-hand panel shows the criteria originally from Lacy et al. (2004), 
  but updated in Lacy et al. (2006). Note that 
  the axes have been selected to facilitate comparison with the original works.
  Crosses indicate typical uncertainties on the mid-IR photometry.  
  HzRGs are the quintessential type 2, luminous AGN.  As expected, they 
  generally reside with the empirical AGN wedges. Tracks from redshift 0 to 
  2 for a late type star-burst galaxy (M82; dashed line) and an early type 
  galaxy (NGC4429; dotted line) are included. These template SEDs are from 
  Devriendt et al. (1999) and circles mark the $z=0$ starting points and 
  asterisks indicate $z =1$ and $z =2$ for the tracks. The tracks indicate 
  that neither galaxy would  be selected as an AGN candidate if located 
  below $z=2$.
  \label{fig.colcol}}
\end{figure}

Our {\it Spitzer} HzRG sample represents the largest sample
of confirmed type 2 quasars with IRAC photometry obtained to
date. Fig.~\ref{fig.colcol} presents the measured mid-IR colors of the
49 HzRGs with detections in all four IRAC bands and the nine HzRGs with
two-band IRAC detections on the color-color plots of \citet{Lacy:04} and
\citet{Stern:05b}. As expected, the HzRG sample tends to fall within the
AGN wedges identified in each of these diagrams.  The \citet{Lacy:04}
region includes nearly the entire HzRG sample: 51/58 (88\%) of HzRGs
plotted match the \citet{Lacy:06} color criteria \citep[which are slightly
updated from the criteria of][]{Lacy:04}.  The smaller \citet{Stern:05b}
region misses a few more HzRG, with only 41/58 HzRGs (71\%) matching the
\citet{Stern:05b} color criteria.  For both plots, nearly every HzRG is
within $1\sigma$ of matching the AGN color criteria.

It should be noted that while the obscured and unobscured AGN generally
lie within the ``AGN wedges'' in Fig.~\ref{fig.colcol}, these two classes 
of luminous
AGN are presumed to have different energetics powering their observed
mid-IR, IRAC-band SEDs.  Luminous, unobscured AGN tend to have power-law
SEDs due to the central engine across the mid-IR regime 
\citep[\eg\,][]{Richards:06}. This leads to rising spectra through the
IRAC passbands and places such sources squarely in the wedges at most
redshifts.  The caveat is that strong features can cause excursions
as they redshift into the passbands; e.g. H$\alpha$ at $z \approx 4$
is in the $3.6\,\mu$m band (e.g., Bian et al., in prep.).  In contrast
to this, we presume that starlight plays a dominant role in the shorter
wavelength IRAC passbands for most obscured AGN, with AGN-heated
dust dominating the reddest IRAC passbands.  Indeed, a clear dip between
the stellar component and the hot dust component is seen in many sources
in our sample.  Consider, as an example, the left-hand panel of Fig.~3.
The redshifted galaxy tracks have [3.6]$-$[4.5]$>0.2$ (Vega) at $z \simgt
1$ due to the $1.6\,\um$ peak of the stellar SED red-shifting into the
$4.5\,\um$ IRAC passband.  The red [3.6]$-$[4.5] colors is then usually
due to the $z > 1$ stellar populations while the red [5.8]$-$[8.0]
colors are due to the AGN.  The next section will concern using
the longer-wavelength {\it Spitzer} data to model and subtract the AGN
contribution to the mid-IR SEDs of our sample, and thus to derive the
rest-frame near-IR stellar luminosities.

\section{Modeling of the Rest-Frame Near-IR and Mid-IR }

We begin our analysis of the broad-band IR SEDs by considering the
galaxies with the most extensive {\it Spitzer} coverage.  This smaller
sample of 26 galaxies illustrates the shape and range of HzRG SEDs in the
IR (\S4.2), and is then used to inform analysis of the sources with more
restricted {\it Spitzer} imaging (\S4.3).  

Despite the extensive near-IR data available for many of the HzRGs,
we elect not to include these data in the SED fitting (though the
$K$-band photometry is over-plotted in the SED plots).  This omission
has several reasons and, as described in \S5, should have minimal
impact on the derived stellar masses, the primary goal of our analysis.
Foremost, we omit the near-IR data as it samples the rest-frame UV to
optical wavelengths of our high-redshift galaxies.  At these wavelengths,
several issues arise.  First, scattered and direct AGN continuum is blue
in color and will thus be a larger contributor to the SED at wavelengths
below 4000\,\AA\ \citep[\eg][]{Vernet:01}.  Second, high equivalent
width emission lines can significantly affect the near-IR photometry for
certain redshifts \citep[\eg][]{VillarMartin:06}.  Indeed, when H$\alpha$
is in the bluest IRAC channels (for our highest redshift sources), we do
not include the affected channel in the SED modeling.  Third, possible
contributions from younger stellar populations will be most pronounced
at shorter wavelengths.  Fourth, rest-frame optical and UV data are more
susceptible to dust extinction than rest-frame near-IR data.  The final
reason to omit the near-IR data is the large inhomogeneity of the data
quality, depth, and analysis techniques.  We are generally relying upon
published photometry done on 3\,m to 10\,m telescopes over the past two
decades and extracted with a range of aperture sizes.  Only in a few
cases are we able to obtain the images from telescope archives.

For the sake of homogeneity of the analysis, the SED modeling described
below relies solely upon {\it Spitzer} photometry.  In a forthcoming
paper, we study the detailed stellar population properties of the HzRG
sample, combining {\it Spitzer} data with archival data and data from
our ongoing observing campaigns at Palomar Observatory and the VLT
(A. Rettura et al., in prep.).  For a few galaxies, we have already
performed the full analysis using all available (optical to mid-IR)
data \citep[\eg][]{VillarMartin:06, Stern:06, DeBreuck:07}.  As shown
in \S5, the derived stellar masses are consistent with the results of
the simpler analysis presented here.

\subsection{The Model}

Twenty-six of our radio galaxies have both IRAC and MIPS observations,
of which only four are not detected at $24\, \mu$m (or in the longer
wavelength MIPS channels). These four
sources include three of the highest redshift radio galaxies and one radio
galaxy at $z \sim 1.5$.  In the latter case, this non-detection may be 
indicative of silicon absorption at rest-frame $9.8\, \um$.  Alternatively, 
the undetected $z \sim 1.5$ radio galaxy, LBDS~53W091, 
is a low luminosity radio source and \citet{Stern:06} argue 
that its mid-IR non-detection likely indicate a SED dominated by old 
stellar populations and not by hot, AGN heated dust. One source, 
6C~0140+326, cannot be distinguished from a bright foreground galaxy 
at 3.6 and $4.5\,\um$ so we do not use it in our modeling here. Hence our 
final sample numbers 21 HzRGs with IRAC and MIPS detections.

For the 21 HzRGs with MIPS detections, we fit the SEDs with toy models
comprised of a Composite Stellar Population (CSP) and multiple black-body 
(BB) dust components.  Our primary goal is to simply derive the rest-frame
near-IR stellar luminosity, which can then be translated into a stellar
mass assuming reasonable stellar population synthesis models (\S5). 
We are also, in these cases with MIPS detections, able to estimate the 
rest-frame $5\,\um$ monochromatic luminosities, a possible measure of the 
obscured AGN power \citep{Ogle:06}.
The difficulty is that in many sources a steeply rising continuum
is visible in the redder IRAC bands, indicative of the tail of hot,
AGN-heated dust contaminating the rest-frame near-IR emission.
Owing to the sparse sampling of the SEDs at these wavelengths, our goal
is not to fully explore the astrophysics of the continuum emission
at these wavelengths, but to use the MIPS (and IRS, where available)
photometry to model the dust SED and subtract its contribution at
rest-frame near-IR wavelengths.  We do not apply Galactic extinction
corrections as they are negligible ($<0.5\%$) at the {\it Spitzer}
wavelengths.  Also, since emission lines longward of H$\alpha$ are not
strong enough to significantly affect our modeling 
\citep[\eg\,][]{Haas:05, Ogle:06,Cleary:07}, they are ignored in the 
analysis. Additional systematic issues in our modeling are discussed in 
\S4.4 and \S6.3, but we stress here that the stellar masses are very 
insensitive to the choice of BBs for the longer wavelength component.

We model the radio galaxy IR SEDs with a four component model consisting
of an elliptical galaxy CSP template and three BBs of different dust
temperatures.  Best-fit parameters for the models are determined using
standard $\chi^2$ minimization techniques. The uncertainties on the
flux densities used to determine the $\chi^2$ of a specific fit include
both random (from the observations) and systematic (\eg\, the detector
zero points and the shape of the SED) uncertainties, which are added in
quadrature (see \S3). The full fitting is only done for sources with 
MIPS $24\um$ detections as we need both enough data points to fit our 
model and a detection longward of the IRAC bands to provide sufficient 
leverage on the dust component (see \S4.2). For sources 
without MIPS detections or observations we do not have enough data points 
to perform a secure fit of this four component model. Hence in many cases 
we only get an upper limit to the stellar component, though in some 
cases we can get a likely value for the stellar luminosity when the IRAC 
bands fit the stellar component well (see \S4.3).

\subsubsection{Stellar Component}

The CSP template used in our SED modeling consists of an elliptical
galaxy modeled with {\tt P\'EGASE.2} \citep{Fioc:97}.  The adopted star
formation history (SFH) scenario has been shown to result in reliable
photometric redshifts with the code {\tt Z-PEG} \citep{LeBorgne:02} and
to well reproduce early-type galaxy SEDs (from rest-frame UV to near-IR)
up to $z \sim 1.25$ \citep{Rettura:06}. Specifically, the star-formation 
rate is proportional to the gas density, $SFR \propto \rho_{\rm gas}$, and the 
astration (recycling) rate is equal to $3.33$ Gyr$^{-1}$.  
Gaseous exchanges with the interstellar medium are represented by 
infall and galactic winds. The former process simulates mass growth 
and the latter represses any star-formation after the first Gyr by 
dissipating the gas from the galaxy and hence driving the gas fraction 
towards zero. The reader is referred to the aforementioned papers for more 
details on the adopted template parameters and the star-formation histories.
Additionally \citet{RoccaVolmerange:04} demonstrates how this elliptical 
galaxy model works for the most massive galaxies that populate the lower 
envelope of the $K-z$ relation. For the current work, we use a \citet{Kroupa:01} 
initial mass function (IMF) and dust-free model templates. We note that in 
the templates used here, the metallicity is not a free parameter; it evolves 
consistently with the SFH, reaching solar values at the ages of interest in 
this study. We assume a galaxy formation redshift of $z_{\rm form} = 10$, 
and we discuss the systematic effects of this choice, along with the other 
details of the galaxy models in \S6.3.  In brief, however, though the 
derived stellar masses vary somewhat with the model choices (mainly by a 
simple offset depending on the choice of IMF), we find that the derived 
luminosities are relatively robust against model uncertainties for stellar 
population ages greater than 1\,Gyr.

\subsubsection{Dust Component}

The longer wavelength modeling assumes three dust components, each
of a different temperature.  The coldest of the BBs is chosen to have
a temperature 60\,K, representative of the large cool dust reservoirs
associated with radio galaxies \citep[\eg][]{Greve:06}, although this
component does not contribute to the rest-frame near-IR.  The second
BB component has a fixed temperature of 250\,K.  For the final, hotter BB
component we allowed the temperature to vary from 500 to 1500\,K (in steps 
of 50\,K) and interpret it as being associated with hot dust heated by the 
central AGN. This upper limit of 1500\,K was chosen as dust sublimates at
this temperature. Systematic effects from the chosen BB temperatures are 
discussed in \S6.3; in brief, we find they have little effect on the 
derived stellar masses. We have also experimented with toroidal AGN models
\citep[\eg][]{Pier:92,Granato:97,Nenkova:02,Siebenmorgen:04}.  However,
the paucity of our infrared data, consisting simply of photometric points
and not infrared spectra, makes fitting with detailed models ill-constrained. 
We emphasize that this dust modeling is not completely 
physical; however, the modeling is sufficient for the primary goal of this 
paper, which is to simply estimate and subtract any contribution by (hot) 
dust to the rest-frame near-IR emission.

\subsection{Fits to HZRGs with MIPS Detections}

The best fit models for the 21 HzRGS with MIPS detections are shown in
Fig.~\ref{fig.mipsfit}.  Observed wavelength is indicated along the bottom
axis and rest wavelength is on the top axis.  Open circles, not used
in the fitting, are either (i) $K$-band data or (ii) potentially
contaminated by strong emission lines (\eg\, H$\alpha$). Upper limits
are indicated by downward arrows. We separate the contribution of host
dust and stellar emission at rest-frame $H-$\,band, which can then be used
to derive stellar masses (\S5).  Table~\ref{table.sedfitting} presents
the total $H-$\,band luminosities, the stellar $H-$\,band luminosities, the
stellar fractions at $H-$\,band, stellar masses, and rest-frame $5\, \um$ 
monochromatic luminosities for sources with full SED fitting.  This fiducial 
mid-IR wavelength, $5\, \um$, was selected as it is always short-ward of the
$24\,\um$ MIPS observations, even for our most distant HzRGs with $24\um$
detections ($z \sim 3.8$).  The formal $\chi^2$ uncertainties ($95.4\%$
confidence limits) of the stellar luminosities from the fitting are
listed in Table~\ref{table.sedfitting}.  They are generally of the
order $\sim25\%$ and are mainly due to the uncertainty in the IRAC
flux densities, which, in turn, are dominated by systematics. 

\subsection{Fits to HzRGs without MIPS Observations or Detections} 

We can constrain, or in some cases quantify, the stellar luminosities for
sources without MIPS observations or detections using the results
of the full SED fitting described above.  For many HzRGs we clearly
observe the $H-$\,band stellar peak (see Fig.~\ref{fig.mipsfit}).
Even in those cases where we do not see the $H-$\,band peak, IRAC channels 
one and two are typically dominated by the stellar component.  At the least, 
we can obtain an upper limit to the stellar mass by demanding that the 
stellar component does not exceed any of the IRAC photometric points.
Hence we can obtain upper limits to the stellar masses of these HzRGs.
Typically, this upper limit is restricted by just IRAC channel 1, 
as the observed mid-IR SEDs are often quite red. In many cases, the 
maximum fit is constrained by IRAC channels 1 and 2, and occasionally by
longer wavelength channels.  In almost half of the radio galaxy sample,
the stellar SED is well fit in two or three channels (those at shorter
wavelength), so we nominally fit the SED and derive stellar luminosities.
The results of the fitting for the HzRGs without MIPS observations or
detections are shown in Fig.~\ref{fig.iracfit}, which adopts the same
symbol convention as Fig.~\ref{fig.mipsfit} bar the inclusion of BB dust
components. As can be seen, the $16\,\um$ observations, where present, 
generally show
a strong excess above the maximally fitted stellar template.  In what
follows, we distinguish those HzRGs whose IRAC SEDs are well fit by a
stellar population (e.g., where multiple IRAC channels are well fit by
the stellar SED model) from those where the IRAC photometry has merely
provided an upper limit to the rest-frame near-IR stellar luminosity.

\subsection{Systematic Effects in the SED Modeling}

We discuss here two possible systematic effects in the SED modeling: the 
adopted BB temperatures and potential photometric offsets in the longer 
wavelength photometry due to strong spectral features redshifting into 
the {\it Spitzer} passbands. The latter effect mimics the effects of 
alternate SEDs with strong absorption and/or emission features. As seen 
below, neither systematic issue strongly affects the derived rest-frame 
near- and mid-IR luminosities.

As the cooler dust components have negligible influence at rest-frame
$H-$\,band, the exact temperatures used are not vital to the final
derived stellar luminosities and masses.  Modeling observations of 3C
radio galaxies obtained with the {\it Infrared Space Observatory},
\citet{RoccaVolmerange:05} find the far-IR SEDs are well fit with
two BB components of temperatures $40 \pm 16$\,K and $340 \pm 50$\,K.
\citet{Archibald:01}, modeling sub-millimeter observations of HzRGs,
likewise adopts an optically-thin, isothermal grey-body of temperature 
40\,K. When we redo our analysis adopting this cooler cold dust component 
rather than the previously adopted 60\,K, the derived rest-frame near/mid-IR 
luminosities change by less than 0.1\%.  Likewise, using the 340\,K warm 
dust component of \citet{RoccaVolmerange:05} instead of the previously 
adopted 250~K, we find that the derived rest-frame near-IR luminosities 
change by less than 0.1\%, although the mid-IR luminosities increase by 
$\sim40\%$.

The fitting algorithm we adopt assumes a very simplistic model of the
long-wavelength HzRG SED. No HzRGs have been observed to have PAH features 
\citep{Ogle:06, Cleary:07}, but some less luminous, radio-quiet type 2 AGN 
have been seen with PAH features \citep{Lacy:07}. Silicon absorption has 
been observed in HzRGs \citep{Ogle:06}, but even with a line width of 
$1.4\,\um$ and $\tau=1.5$ we would only observe a $30-40\%$ decrease in the 
observed flux density. \citet{Ogle:06} also note that there are indications 
that the power-law continuum is truncated at the sublimation temperature. 

We have investigated the effects of these potential features by re-running 
the fitting procedure, varying the $16\um$ and $24\um$ flux densities to 
simulate those passbands falling on strong emission or absorption features.  
We varied the $16\um$ flux densities by $10\%$ to account for weaker PAH 
features at $6.4\um$ passing through the IRS bandpass.  We varied the 
$24\um$ flux densities by $50\%$ to account for both the strong $7.7\um$ 
PAH emission feature and the strong $9.8\um$ silicate absorption feature.  
These experiments only changed the mean $5\,\um$ luminosities by $\sim5\%$ 
(for varying the $16\,\um$ flux density) and $\sim25\%$ (for varying the 
$24\,\um$ flux density).  Given the lack of observational evidence for these 
features so far we consider these estimates conservative. The effects on the 
mean stellar luminosity (and hence mass) are $\simlt2\%$ in both cases.

\section{Results of SED fitting}

\subsection{Stellar Fractions}

Figures ~\ref{fig.stelfrac} and ~\ref{fig.stelfrac2} show the fraction of 
the HzRG rest-frame $H-$\,band luminosity which is modeled as stellar in 
origin, plotted against total $H-$\,band luminosity, stellar $H-$\,band 
luminosity, redshift, radio power, and rest-frame $5\,\um$ luminosity 
The stellar fraction is defined as the ratio of the flux densities found 
from convolving the stellar and total SEDs with a $H-$\,band filter curve.

Only the 21 HzRGs with MIPS detections are presented in these figures and 
their $H-$\,band luminosities (stellar and total), stellar fractions, 
inferred stellar masses (see \S6), and monochromatic $5\,\um$ luminosities 
(see \S5.4) in Table~\ref{table.sedfitting}.  As can be seen, half the 
sample has the rest-frame near-IR luminosity strongly dominated by 
starlight, with $\simgt 95\%$ of the emission apparently stellar in origin.  
Another $20\%$ of the sample have stellar emission accounting for 
$60 - 90\%$ of the rest-frame near-IR flux, and thus still dominating 
the emission at this wavelength.  There appears to be a weak inverse 
correlation between stellar fraction and {\em total} $H-$\,band luminosity 
whilst the stellar fraction does not vary with {\em stellar} $H-$\,band 
luminosity. This suggests that the underlying host galaxy population is homogeneous. 
No evolutionary trends in the stellar fraction are evident, nor 
does the stellar fraction appear to correlate with either radio power or 
rest-frame mid-IR luminosity. This result is expected for the following 
reasons.  The radio emission at rest-frame 3\,GHz is dominated by isotropic 
lobe emission.  Likewise, rest-frame mid-IR samples largely isotropic 
emission coming from near the central engine \citep[\eg\,][]{Heckman:94}; at 
these wavelengths, the emission should be largely immune to extinction. The 
stellar fraction at $H-$\,band, however, is very dependent upon obscuration of 
the AGN which, in the currently preferred unified schemes 
\citep[\eg\,][]{Antonucci:93, Urry:95}, is a function of orientation.
AGN with little obscuration of the central engine will have higher
rest-frame near-IR luminosities, show broad lines, and have low stellar
fractions.  Indeed, the three sources in our MIPS sample with the highest
rest-frame $H-$\,band luminosities, PKS~1138-262, TX~J1908$+$7220 and B3~J2330$+$3927,
both show AGN signatures in the near-IR (see Appendix) and are indicated as 
upper limits in Table~\ref{table.sedfitting} and the relevant figures 
(Figs.~\ref{fig.stelfrac}, \ref{fig.stelfrac2}). The stellar fraction is expected 
to be an indicator of orientation and should thus not correlate with the 
orientation-independent parameters. In a future publication (Seymour et al. 
in prep.), we shall analyze how the stellar fraction correlates with
properties which {\em are} expected to depend on orientation, such as
radio core fraction and radio spatial extent.

\subsection{Rest-frame Near-IR Stellar Luminosities}

The rest-frame $H-$\,band stellar luminosities are found mainly to lie in
the $10^{11}-10^{12}L_\odot$ range, as illustrated in Fig.~\ref{fig.logh}.
Overlaid are P\'EGASE models illustrating the luminosity of an elliptical 
galaxy model with $z_{\rm form}=10$ for masses of $10^{11}$ and
$10^{12} M_\odot$.  The derived HzRG stellar luminosities are consistent
with extremely massive hosts, with masses of a few $\times 10^{11}
M_\odot$ out to $z = 4$. Derivations of masses and their associated 
uncertainties are presented in \S6.

This result confirms earlier work on the masses of radio galaxies. 
\citet{RoccaVolmerange:04} find that most radio galaxies have $K-$\,band
magnitudes consistent with stellar masses of $10^{11}-10^{12}M_\odot$
with a few having magnitudes implying masses $>10^{12}M_\odot$. These 
sources with bright $K-$\,band magnitudes may have emission lines or 
scattered AGN light contributing to them so their $K-$\,band should be 
viewed as upper limits.

We also plot the rest $H-$band stellar luminosities of a sample of sub-mm 
galaxies (SMGs) in the GOODS North region from \citet{Borys:05} as red
crosses in Fig.~\ref{fig.logh}. The $H-$\,band luminosities are re-derived 
using our modeling technique with the data presented in \citet{Borys:05}.  
Our analysis produces similar luminosities to \citet{Borys:05}. However, 
due to different assumed star-formation histories and IMFs, our derived 
stellar masses systematically differ by $25\%$. These SMGs have a similar range of 
rest-frame $H-$\,band luminosities to the HzRGs, implying similar stellar 
masses. There is a probable evolutionary link between SMGs and HzRGs, 
though the number of similarities between these populations is comparable 
to the number of differences. On the one hand, SMGs have a much higher 
areal density than HzRGs:  the former are relatively numerous in 
small-area (100 arcmin$^2$) surveys \citep[\eg][]{Borys:03}, while 
the latter are much rarer and typically herald from large-area or 
full-sky radio surveys. On the other hand, many radio galaxies are 
strong sub-mm emitters \citep[\eg][]{Archibald:01, Greve:06} and would 
qualify as SMGs irrespective of their strong radio emission.  

\subsection{Mid-IR Luminosities}

The rest-frame $5\,\um$ monochromatic luminosities presented in
Table~\ref{table.sedfitting} are generally all above $10^{11}\,L_\odot$.
In Fig.~\ref{fig.z5um} we show the distribution of the $5\,\um$
luminosities with redshift.  The solid symbols and arrows indicate the
21 HzRGs with MIPS detections and the five HzRGs with MIPS upper limits.
The open symbols indicate the 13 HzRGs with IRS observations but no MIPS
observations where the IRS band is close to rest-frame $5\,\um$.
These IRS-derived luminosities are calculated directly from the $16\,\um$
flux densities, and thus do not account for the shape of the SED or the
related offset between rest-frame $5\um$ and the rest-frame wavelength
probed by the $16\um$ observations.  The uncertainty on the IRS-derived
rest-frame $5\um$ luminosities is estimated to be $\simlt 50\%$.

The high-luminosities, $L_{5\um}\simgt10^{11}\,L_\odot$, in 
Fig.~\ref{fig.z5um} imply that the HzRGs are re-radiating a substantial 
amount of energy, most likely re-processed emission from the obscured 
AGN heating the surrounding dust. The implied total ($5-1000\um$) IR 
luminosities will be greater than the mid-IR luminosities presented 
here and hence most of these HzRGs would be classified as LIRGs 
($L_{5-1000\um}>10^{11}\,L_\odot$), or in many cases as ULIRGs
($L_{5-1000\um}>10^{12}\,L_\odot$) had they been initially identified
with no knowledge of their radio properties.  These luminosities
cover a similar range as found by \citet{Ogle:06} for a sample
of quasars and radio galaxies from the 3CRR catalogue:  $\nu
L_\nu$(rest $15\,\um$) $=10^{44-46}\,\ergps$.  Luminosities this
high confirm that powerful radio galaxies harbor ``hidden quasars'',
\ie\, super-massive black holes accreting near the Eddington limit.
A few sources have lower luminosities, classified by \citet{Ogle:06}
as being in the mid-IR ``weak'' regime, $\nu L_\nu$(rest $15\,\um$)
$<10^{44}\,\ergps$, approximately equivalent to $\nu L_\nu$(rest $5\,\um$)
$<10^{11}\,L_\odot$, indicating that they probably have lower accretion
rates, and are possibly non-thermal, jet-dominated sources.

The alternate explanation for these high mid-IR luminosities, 
star-formation, is unlikely. Despite the high inferred star-formation 
rates from the sub-mm observations \citep{Archibald:01, Reuland:03}, 
$\ge500M_\odot$yr$^{-1}$ for the $\sim30\%$ detected and 
$\le500M_\odot$yr$^{-1}$ for the non-detections, the star-formation 
is widely scattered as seen in the rest-frame UV/optical images
which show widely distributed morphologies 
\citep{McCarthy:87,Best:96,Zirm:03,Miley:06}. Hence the star-formation
is unlikely to be dense enough to reach temperatures where it would
dominate at rest-frame $5\,\um$. 

\subsection{Near- to Mid-IR Luminosity Ratio}

Figure ~\ref{fig.ratiof} presents the stellar near-IR to total mid-IR 
($L_{\rm H}^{\rm stel}/L_{5\,\um}^{\rm tot}$) rest-frame luminosity ratio 
plotted against both redshift and $3\,$GHz luminosity. The ratios range from 
$\sim 0.1$ to $\sim 6$ and we 
find that the near- to mid-IR luminosity ratio gradually decreases with both 
increasing redshift and with increasing radio luminosity.  If, as discussed 
earlier, the mid-IR luminosities are indicators of the power of the AGN 
(\ie\, the AGN accretion rate at the epoch of observation), then for a given 
host stellar mass, the accretion rate apparently declines at lower redshift 
and lower radio power.  This correlation suggests evolution with redshift,
but may also be due to selection:  our mid-IR luminosities are based
on the $24\,\um$ detections which are fewer at higher redshift. The
trend with radio power is perhaps more interesting, showing that HzRGs
with more luminous radio emission are also accreting at a higher rates
(for a given stellar mass). This result is also consistent with the radio 
luminosity - emission line luminosity correlation \citep[][]{Willott:99}. 
This trend with redshift is probably also related to the slight 
trend, albeit with a large scatter, when directly comparing the mid-IR to 
radio luminosity (see Fig.~\ref{fig.mir_rad}), although that trend is 
possibly due to the scaling of luminosities with distance.

\section{Photometric Stellar Masses} 

The stellar luminosities shown in Fig.~\ref{fig.logh} are converted to
masses using the mass-to-light ratios\footnote{The modeled mass-to-light
  ratios include the mass of {\em all} material that has gone through
  star-formation.  It thus represents the integrated star formation history
  and includes stars, as well as sub-stellar objects and compact post-stellar
  relics such as white dwarfs, neutron stars, and black holes, but {\em not}
  the gas that is recycled to the ISM. Stars dominate the rest-frame near-IR 
  SED.} derived from the {\tt P\'EGASE.2} early-type galaxy model 
with $z_{\rm form}=10$ described in \S4.1 and are presented in 
Fig.~\ref{fig.logm}. The plotted errors are from the SED fitting of the 
luminosity; systematic errors to the stellar masses are 
discussed in \S6.3. Hence we show that the stellar masses of the hosts of radio 
galaxies generally have masses distributed between $10^{11}-10^{11.5}M_\odot$ 
with remarkable little scatter across the redshift range of $1\le z\le4$. 
The one HzRG with larger uncertainty at $z=5.2$ suggests this trend may hold to 
higher redshifts still with the implication that this source had to have formed
early and quickly. Given the lifetimes of radio galaxies are short, $10-100\,$Myrs,
these sources are likely to be representative of a larger population of 
massive galaxies at these redshifts. Massive galaxies are now being found in 
great numbers across $1\le z\le 3$ \citep[\eg][]{Labbe:05,Papovich:06,vanDokkum:06}.

We compare our HzRG sample to SMGs studied by \citet{Borys:05} (red crosses in 
fig.~\ref{fig.logm}). HzRGs and 
SMGs are amongst the most massive objects over this redshift range, showing 
that both classes of galaxy probe the most massive galaxies in the 
high-redshift universe, though we note that HzRGs are much rarer 
beasts. \citet{Borys:05} studies 13 SMGs identified in the GOODS survey.  
For comparison, the most luminous radio source known in the GOODS survey is
VLA~J123642+621331, a $470 \mu$Jy source at $z = 4.24$ identified by
\citep{Waddington:99}.  With $L_{\rm 3 GHz} \sim 10^{25}\, {\rm W}\,
{\rm Hz}^{-1}$, the radio emission from this galaxy is most likely coming
from an AGN, not star formation, as suggested by European VLBI Network 
observations \citep{Garrett:01} which show the source to be unresolved.  
Nevertheless, it's radio power does not qualify it as a HzRG according to 
our selection criteria defined in \S2.

\subsection{Radio Power Dependence on Stellar Mass}

Fig.~\ref{fig.radlogm} presents the derived stellar masses plotted against 
rest-frame 3\,GHz radio luminosity.  The small dynamic range of the stellar 
masses combined with their relatively large uncertainties makes the formal 
determination  of any correlation difficult, but a weak correlation is 
visually apparent. A proper statistical survival analysis that accounts for 
the multiple upper-limits finds that the Cox Hazard Probability is only 0.5 
meaning there is only a 50\% chance that stellar mass correlates
with radio luminosity for our sample.  The detection of a correlation,
showing that more powerful radio sources are hosted by more massive
galaxies, would not be surprising, and is in fact expected if the $M
- \sigma$ relation holds to high redshift and all HzRGS
have similar efficiency rates or Eddington ratios.  Previous work, inferring
stellar masses from the {\em observed} near-IR flux rather than the {\em
rest-frame} near-IR luminosity, has suggested this correlation in the past
\citep{Eales:97, Best:98, Jarvis:01, DeBreuck:02, Willott:03a}.

\subsection{Comparisons with More Detailed Modeling}

As mentioned earlier, several HzRGs from our sample have already been the
subject of detailed studies that included {\it Spitzer} observations and
determined host stellar masses.  These studies have generally used more 
extensive photometry for the modeling, including optical and near-IR data,
and have used more complicated models to determine the host stellar masses.  
We review the results of more detailed modeling here; in brief, allowing 
for different choices of IMF, the simplistic modeling presented here is
consistent with the more detailed modeling

MRC~2104-242 ($z=2.491$) was analyzed in detail by \citet{VillarMartin:06}
who found a stellar mass of $5\pm2\times10^{11}M_\odot$, $E(B-V)=0.4$,
and an age of 1.8\,Gyr. This result is consistent with the work here where we
find a mass of $1.6^{+1.6}_{-0.8}\times10^{11}M_\odot$.  Our work assumes
no extinction and $z_{\rm form} = 10$ corresponds to an age of 2.1\,Gyr
for the redshift of this HzRG.  Much of the difference can be attributed
to the \citet{Salpeter:55} IMF adopted by \citet{VillarMartin:06}
which yields a mass-to-light ratio systematically higher by a factor of
$\sim 40\%$ compared to the \citet{Kroupa:01} IMF adopted
here.  Taking this choice into account, the derived masses are within $1
\sigma$ of each other, which is rather encouraging given the increased
sophistication of the stellar modeling in \citet{VillarMartin:06},
as well as their inclusion of near-IR data from the ground and {\it HST}.

LBDS~53W091 ($z = 1.552$) was studied by
\citet{Stern:06} and, as discussed earlier, is (proto-)typical of the old,
evolved class of extremely red objects.   Using the \citet{Salpeter:55}
IMF, \citet{Stern:06} find a stellar mass of $\sim3\times10^{11}M_\odot$, 
and an age of $3.5 - 4$\,Gyr.  Accounting for the difference
in IMF, this result is consistent with the masses presented here,
$\sim1.6\times10^{11}M_\odot$.  The ages of the LBDS galaxies imply a
redshift of formation $z_{\rm form} \sim 9$, similar to our $z_{\rm form}=10$.

The full radio to X-ray SED of 4C~23.56 ($z = 2.483$) is presented
in \citet{DeBreuck:07}, the first time such data has been presented for a 
radio galaxy above $z=2$.  A full SED decomposition of the AGN and stellar 
emission (such as scatter quasar light, emission lines and nebular 
continuum) is performed. With a Salpeter IMF and no extinction, they derive 
a stellar mass of $5.1^{+1.6}_{-2.1}\times10^{11}M_\odot$.  Again, allowing
for the different choice of IMF, this result is consistent within $1\sigma$ 
of the results here, $3.9^{+1.2}_{-0.7}\times10^{11}M_\odot$.  

These comparisons give us confidence that though the
SED modeling we adopt is somewhat simplistic, the derived stellar masses
appear to be accurate within the uncertainties inherent to the enterprise.

\subsection{Systematic Effects on the Stellar Masses}

Deriving stellar masses of high-redshift galaxies ($z \simgt 1$) has
become widespread in recent years with the advent of deep near- and
mid-IR extra-galactic surveys \citep[\eg\,][]{Shapley:03, Labbe:05,
Rettura:06, Moustakas:07}.  {\em Photometric} masses, as derived here, are
subject to several systematic uncertainties.  However, ``relative''
masses can be derived with less uncertainty \citep[\eg\,][]{Dickinson:03}.
We now briefly discuss some of the systematic effects inherent to the
stellar masses derived here.  This analysis will be particularly important 
for future comparison of our results with other samples.

\subsubsection{IMF}

The IMF remains one of the largest uncertainties when modeling the
SED of a galaxy. There have been several updates to the classical
\citet{Salpeter:55} IMF recently \citep[\eg\,][etc.]{Kroupa:01,
Chabrier:03}.  Even these, however, are largely based on local stars
within our own Galaxy and hence suffer uncertainties at the low mass end
and for extreme metalicities (high and low).  We have chosen to use the
\citet{Kroupa:01} IMF here.  However, our results would not be very
different ($<2\%$) if we used the \citet{Chabrier:03} IMF. Had
we used the classical \citet{Salpeter:55} IMF, our stellar masses would
increase systematically by $40\%$, independent of our choice 
of $z_{\rm form}$.

\subsubsection{Formation Redshift}

We have assumed a uniform formation redshift, $z_{\rm form}=10$, in the 
current analysis. \citet{Eyles:06} find a range of $z_{\rm form}=7-14$
for $i'$-band drop-outs at $z\sim6$ in the GOODS fields for galaxies. These
distant galaxies are our best-studied examples of galaxies at $6 \simlt z
\simlt 7$ and hence are our best (only) candidates for the parent
populations to HzRGs. This range of formation redshifts corresponds to 
$0.16\,$Gyr earlier or $0.3\,$Gyr later than $z=10$
when the universe was only $\sim1\,$Gyr old. Assuming the full range
of $z_{\rm form}$ from \citet{Eyles:06} changes the mass by $\pm17\%$
for a \citet{Kroupa:01} IMF and $\pm8\%$ for a \citet{Salpeter:55} 
IMF. The change is such that the stellar mass decreases with a later 
formation redshift and vice-versa.

\subsubsection{Extinction}

No internal galactic extinction is considered here.  However, $\sim50\%$ of
the HzRGs have sub-mm detections, implying substantial dust content, though
the dust distribution need not heavily obscure
the old stellar population observed at rest-frame near-IR wavelengths.
To obtain a $20\%$ under-estimation of the stellar mass from the 
rest-frame $H-$\,band luminosity requires an $A_{\rm V}>1.17$ assuming a 
\citet{Calzetti:94} extinction curve.

\subsubsection{Star-burst Component}

These stellar mass estimates do not include the possible contribution of 
young stellar population which would lead to an over estimation of the 
stellar mass. If we include a $10\%$ by mass starburst component, starting
200\,Myr before the observed redshift, we find that while the SED at 
rest-frame near-IR does not change significantly, the mass-to-light ratio 
does decrease by $10-20\%$ at $H-$\,band (see Fig.~\ref{fig.m2l}). Redoing 
our analysis again with this younger component results in nearly identical 
$H-$\,band luminosities as before, but a $\la20\%$ decrease in stellar mass. 
We note that despite several sources in our sample having sub-mm detections, 
it is not thought that on-going star formation constitutes as much as $10\%$ 
by mass for such systems.

Deep rest-frame UV spectropolarimetry of nine optically bright star-forming 
HzRGs by \citet{Vernet:01} finds only one object with a substantial contribution 
from a population of young stars. Hence we conclude that there may be 
a mild systematic over-estimation of the stellar mass, but this uncertainty is 
likely less significant than other uncertainties discussed above, such as 
the choice of the IMF in the modeling.

\subsubsection{TP-AGB Stars}

\citet{Maraston:05} has argued that the thermally pulsing asymptotic
giant branch (TP-AGB) phase in the evolution of a simple stellar
population is not properly considered in other spectral evolution
synthesis codes \citep[\eg\, {\tt P\'EGASE} and][]{Bruzual:03}, and
hence analysis using these models over-predicts the near-IR mass-to-light
ratio for galaxy ages $\simlt 1$\,Gyr.  For our choice of $z_{\rm form}$,
this age corresponds to galaxies at $3 \simlt z \simlt 4$.  We emphasize that
the {\tt P\'EGASE} models {\em do} incorporate the TP-AGB phase, just
in a different manner than the \citet{Maraston:05} models.  If this TP-AGB 
phase is indeed an issue, it is most likely to only affect the highest 
redshift (and hence youngest) sources in our sample by decreasing their
stellar mass by $\le30\%$. As we have seen in 
\S6.3.4 an un-accounted star-burst leads to an overestimation of the 
stellar mass. Use of the model from \citet{Maraston:05} would hence 
increase this overestimate to about $30\%$ which is on par with the 
uncertainty due to all the other systematics.

\section{Discussion and Conclusions}

We have performed a comprehensive infrared survey of 69 HzRGs above $z=1$,
the first time such galaxies have been detected in the rest-frame near-IR
and mid-IR. This work presents the largest sample of confirmed type 2
quasars to be systematically studied with {\it Spitzer}.  As expected, they 
predominantly reside in the IRAC color-color space dominated by type 1 AGN
found empirically by \citet{Lacy:04} and \citet{Stern:05b} and theoretically 
by \citet{Sajina:05}.

We model the rest-frame near to mid-IR SEDs derived from our broad band IRAC, 
IRS and MIPS imaging. We use a toy model comprised of thermal BBs to represent 
the emission at longer wavelengths presumably due to dust, and the P\'EGASE
code to model the stellar component. Our analysis yields total near-IR 
($H-$\,band) and mid-IR ($5\,\um$) rest-frame luminosities and we deconvolve 
the stellar and AGN contributions at $H-$\,band. The scatter of stellar 
$H-$\,band luminosities is lower than that for the total 
$H-$\,band luminosities and we find the HzRGs with lower stellar 
fractions tend to have higher rest-frame $H-$\,band total luminosities, 
although this trend is weak. When $H-$\,band luminosities are corrected for AGN
contribution they show no correlation with stellar fraction. Also the 
stellar fraction is not found to correlate with redshift, radio luminosity 
or mid-IR luminosity. These results are consistent with orientation-driven 
models of AGN unification. 

We find that the $H-$\,band stellar luminosities are consistent with stellar
masses of $10^{11-12}M_\odot$, comparable with the masses of local massive
radio galaxies \citep{McCarthy:93, Jarvis:01, Willott:03a}. This conversion 
of luminosity to mass assumes a dust-free, passively evolving elliptical 
formed at $z=10$ with no recent episodes of star-formation. We discuss 
these effects and other systematics and conclude that our conclusions are 
robust at the level of $\approx 10 - 20\%$.  We also note that effects can 
act to either increase the inferred mass (the presence of dust) or to decrease 
it (recent star-burst activity). So multiple effects may cancel out, thus, 
most likely, simply increasing the scatter in the estimated stellar masses.

The derived stellar masses average $10^{11.4}M_\odot$ across $z=1-4$ with 
a scatter less than $50\%$. This scatter is remarkably small, suggesting 
that the tightness of the observed $K-z$ relation at $z>1$ is due to the 
intrinsic homogeneity of the HzRG population. The increase in the scatter 
from $z<1$ to $z>1$ can largely be attributed to the $K$-band being affected 
by emission lines, scattered light from the AGN \citep[\eg][]{DeBreuck:07}
and, potentially, young stellar populations. The scatter in the observed 
$H-$\,band luminosities can be explained by varying contributions from AGN 
heated dust entering the IRAC bands.

These masses, while very high, are not uncommon at these redshifts. Much 
work has been done recently to detect massive, $M>10^{11}M_\odot$, galaxies
at high redshifts \citep[\eg][]{Labbe:05,Papovich:06,vanDokkum:06}. Densities
of approximately one $M>10^{11}M_\odot$ galaxy per square arc-minute are found 
in the redshift range $2\simlt z \simlt 3$ \citep{Papovich:06,vanDokkum:06}. 
Our survey targeted a sample of known sources from the literature,
derived from multiple radio surveys.  As such, we are ill-posed to address
space densities from the current work.  However, we note that our adopted
definition of a HzRG ($z > 1$ and $L_{\rm 3 GHz} > 10^{26}\, {\rm W}\,
{\rm Hz}^{-1}$) implies that large area surveys such as the Faint Images
of the Radio Sky at Twenty-centimeters \citep[FIRST;][]{Becker:95} have
already detected all HzRGs at $1 < z \simlt 3$, though the vast majority lack
spectroscopic observations thus far.  We can therefore constrain the
surface density of HzRGs to be less than 100 per square degree, which is
significantly lower than the surface density of optically and near-IR
selected massive galaxies.  This low surface density is most likely due 
to the short lifetimes ($10-100$\,Myrs) and low duty cycles of the most
powerful radio sources, even if the every massive galaxy were to go 
through a radio-luminous phase.

We find a slight trend that the galaxies with the higher stellar masses
host the more powerful radio sources, a trend suggested by other authors
\citep{McCarthy:93, Jarvis:01, Willott:03a}. However, a full survival analysis 
taking proper account of upper limits finds this correlation has only a 
50\% of being correct.  The correlation is expected if the $M - \sigma$ 
relation holds to high redshift and HzRGs all accrete at comparable 
Eddington ratios.

Most of the HzRG mid-IR luminosities are above $10^{11}L_\odot$, implying
even higher total IR luminosities. These LIRG and ULIRG-like luminosities 
are due to reprocessed radiation from gas and dust obscuring a highly 
accreting AGN. Rest-frame mid-IR luminosity ($\lambda_0 \sim 5 \mu$m) may 
be a good, independent indicator of AGN activity.  Comparison with traditional
AGN indicators at other wavelengths is currently underway (X-ray ---
D. Alexander et al., in prep.; radio --- N. Seymour et al., in prep.).

\acknowledgments
We would like to thank the referee for the constructive comments provided 
on the clarification of this paper.
NS and DS thank ESO for generous hospitality on several 
occasions over the last few years. We thank Dave Frayer and Ranga-Ram Chary 
for advice on reducing the MIPS data and the IRAC instrument team for allowing
publication of GTO data on 4C~41.17 and 6C~0140$+$326. This work is based
on observations made with the {\it Spitzer Space Telescope}, which
is operated by the Jet Propulsion Laboratory, California Institute of
Technology under a contract with NASA. Support for this work was provided
by NASA through an award issued by JPL/Caltech.
The work by WvB was performed under the auspices of the U.S. Department of 
Energy, National Nuclear Security Administration by the University of 
California, Lawrence Livermore National Laboratory under contract No. 
W-7405-Eng-48. WvB also acknowledges support for radio galaxy studies at 
UC Merced, including the work reported here, with the {\it Spitzer Space 
Telescope} via NASA grants SST 1264353, 1265551, 1279182 and 1281587.

\appendix

\section{Notes on Individual Sources}

\begin{figure}
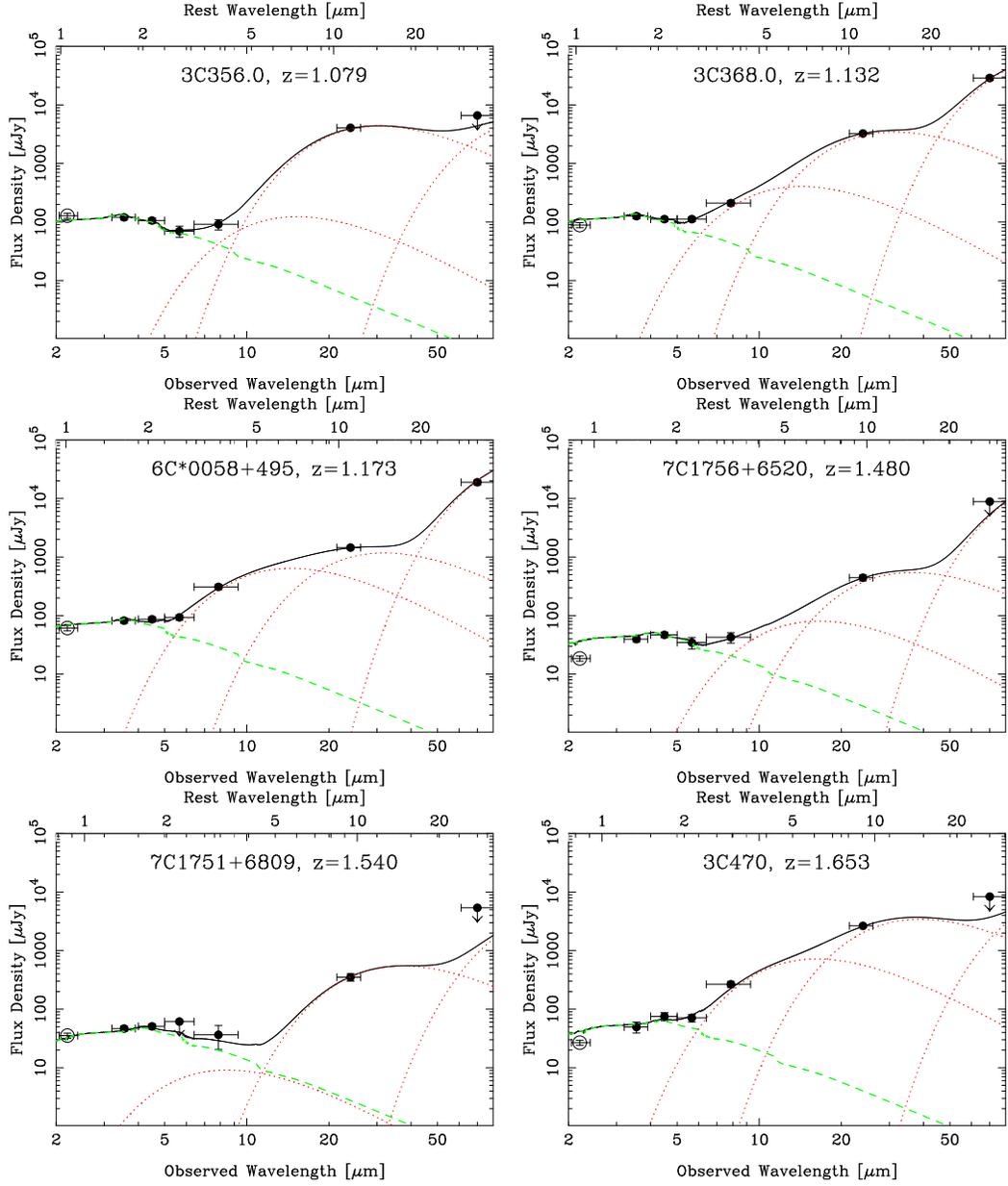

  \begin{center}
  \begin{minipage}{0.33\linewidth}
    \includegraphics[width=5.41cm,angle=270]{f5a.ps}
  \end{minipage}\hspace*{15mm}
  \begin{minipage}{0.33\linewidth}
    \includegraphics[width=5.41cm,angle=270]{f5b.ps}
  \end{minipage}\\
  \begin{minipage}{0.33\linewidth}
    \includegraphics[width=5.41cm,angle=270]{f5c.ps}
  \end{minipage}\hspace*{15mm}
  \begin{minipage}{0.33\linewidth}
    \includegraphics[width=5.41cm,angle=270]{f5d.ps}
  \end{minipage}\\
  \begin{minipage}{0.33\linewidth}
    \includegraphics[width=5.41cm,angle=270]{f5e.ps}
  \end{minipage}\hspace*{15mm}
  \begin{minipage}{0.33\linewidth}
    \includegraphics[width=5.41cm,angle=270]{f5f.ps}
  \end{minipage}\\
  \end{center}
\caption{SED fitting of 21 sources with both IRAC observations and MIPS 
  detections. The name of each HzRG and its redshift are marked at the top of 
  the panel. Observed wavelength is along the bottom axis and rest-frame 
  wavelength is along the top axis. Data with error-bars are presented as 
  filled circles: open circles are not used in the fitting for reasons 
  described in the text. Downward arrows indicate upper limits. The solid 
  dark line indicates the total best-fit SED. The stellar and dust components 
  are indicated by dashed and dotted lines respectively.  
\label{fig.mipsfit}}
\end{figure}

\begin{figure}
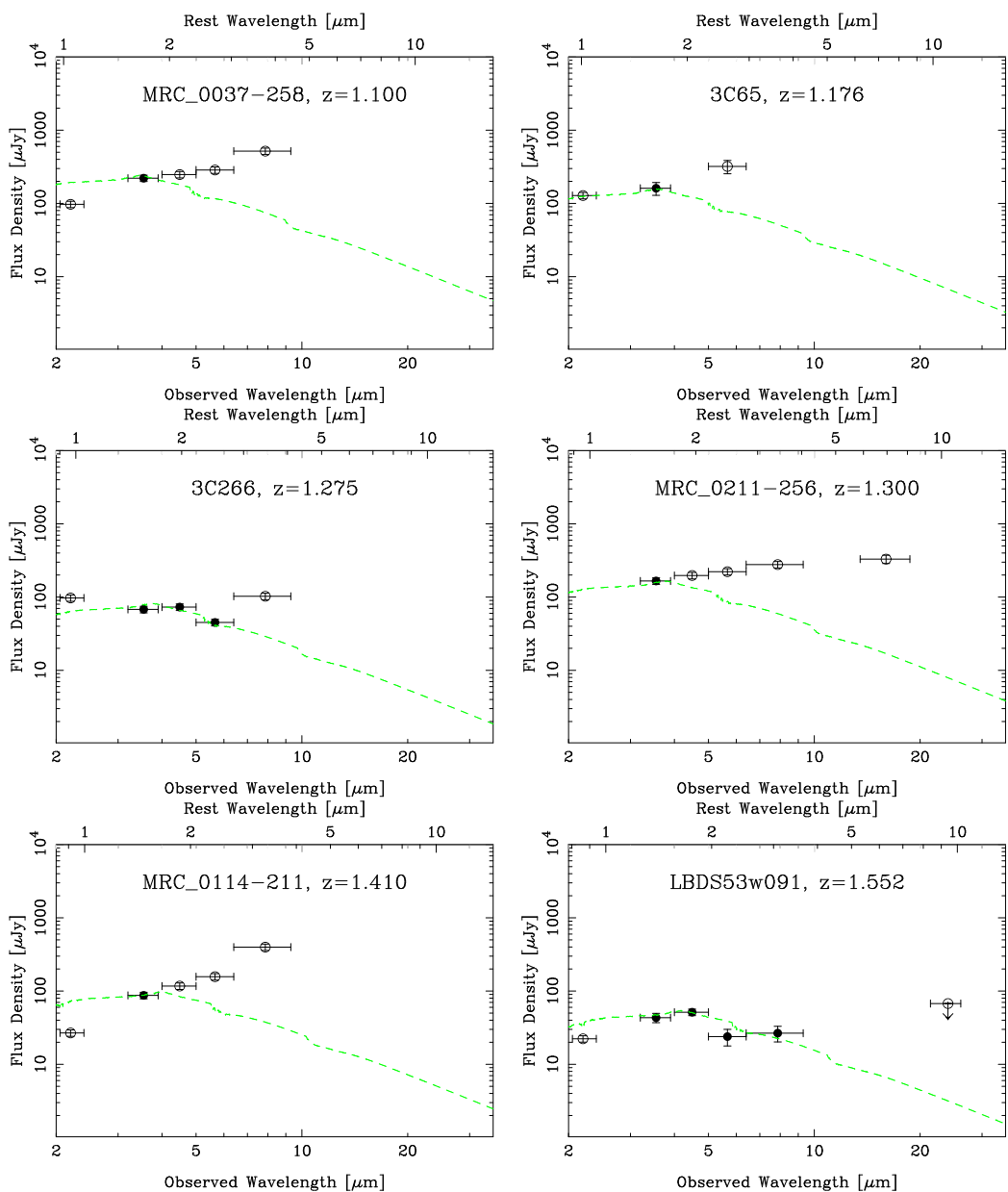

  \begin{center}
  \begin{minipage}{0.33\linewidth}
    \includegraphics[width=5.41cm,angle=270]{f6a.ps}
  \end{minipage}\hspace*{15mm}
  \begin{minipage}{0.33\linewidth}
    \includegraphics[width=5.41cm,angle=270]{f6b.ps}
  \end{minipage}\\
  \begin{minipage}{0.33\linewidth}
    \includegraphics[width=5.41cm,angle=270]{f6c.ps}
  \end{minipage}\hspace*{15mm}
  \begin{minipage}{0.33\linewidth}
    \includegraphics[width=5.41cm,angle=270]{f6d.ps}
  \end{minipage}\\
  \begin{minipage}{0.33\linewidth}
    \includegraphics[width=5.41cm,angle=270]{f6e.ps}
  \end{minipage}\hspace*{15mm}
  \begin{minipage}{0.33\linewidth}
    \includegraphics[width=5.41cm,angle=270]{f6f.ps}
  \end{minipage}\\
  \end{center}
\caption{Galaxy SED fits to determine stellar masses or upper-limits. The 
  name of each HzRG and its redshift are marked at the top of the panel. 
  Observed wavelength is along the bottom axis and rest-frame wavelength 
  is along the top axis. Data with error-bars are presented as filled 
  circles; open circles not used in the fitting are described in the text. 
  Downward arrows indicate upper limits. The dashed line indicates the total 
  best-fit SED. 
\label{fig.iracfit}}
\end{figure}

\begin{figure}
\includegraphics[width=12cm,angle=270]{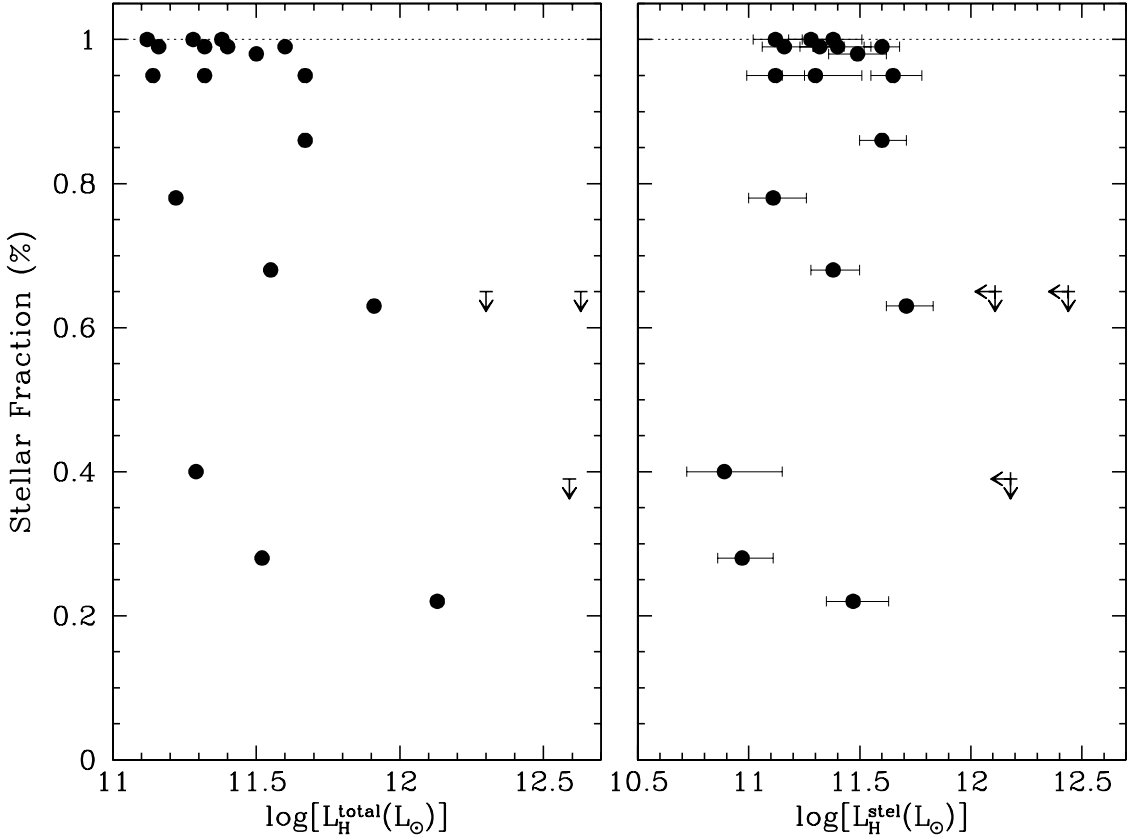}
\caption{Fraction of the HzRG rest-frame $H-$\,band luminosity which is modeled 
  as stellar, plotted as a function of total $H-$\,band luminosity and stellar 
  $H-$\,band luminosity. The stellar fraction appears independent of the stellar 
  near-IR luminosity but is inversely related to the total near-IR luminosity. 
  This implies that the underlying hosts constitute a homogeneous, similar 
  stellar mass population, but with varying AGN contributions to their observed 
  total near-IR luminosities.
  \label{fig.stelfrac}}
\end{figure}

\begin{figure}
\includegraphics[width=12cm,angle=270]{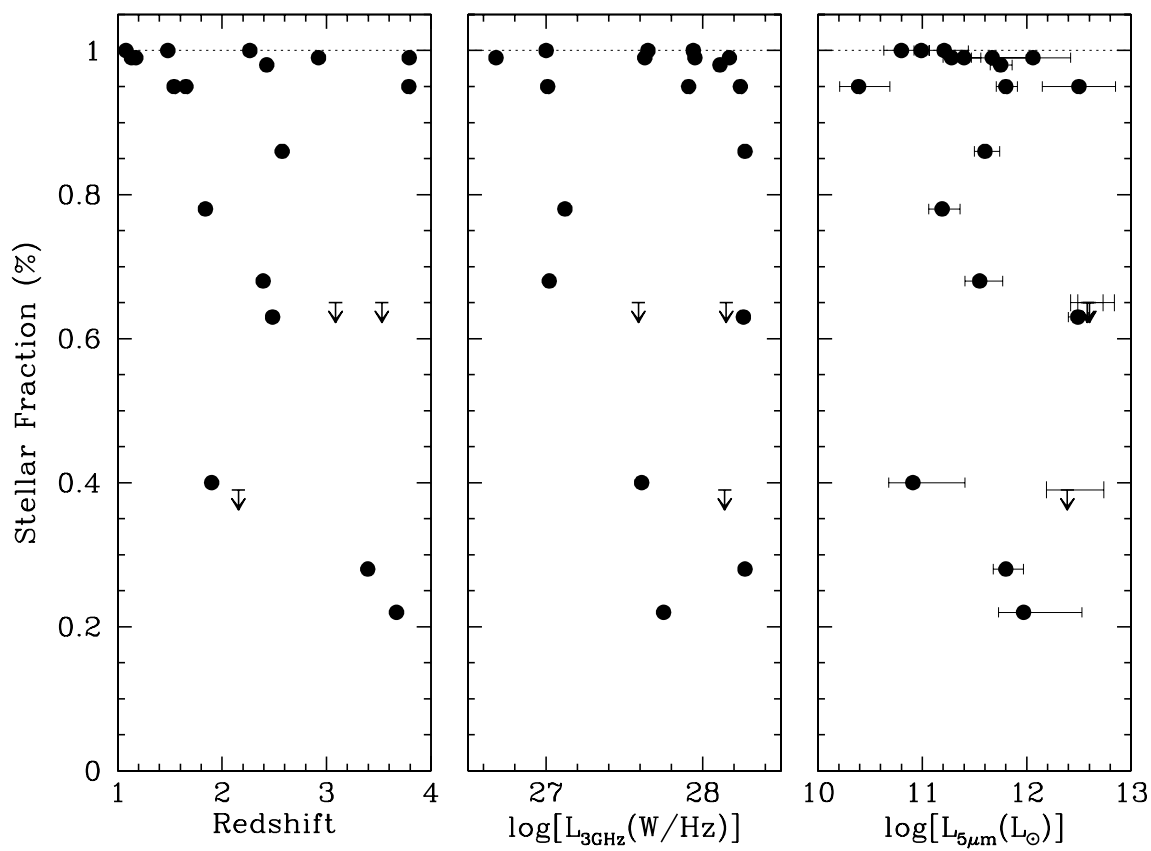}
\caption{Fraction of the HzRG rest-frame $H-$\,band luminosity which is
  modeled as stellar, plotted as a function of (i) redshift, (ii) rest-frame
  3\,GHz radio power, and (iii) rest-frame $5 \mu$m luminosity.  Only the
  21 HzRGs with MIPS detections are plotted.  No correlations are evident. 
  \label{fig.stelfrac2}}
\end{figure}

\begin{figure}
\includegraphics[width=11cm,angle=270]{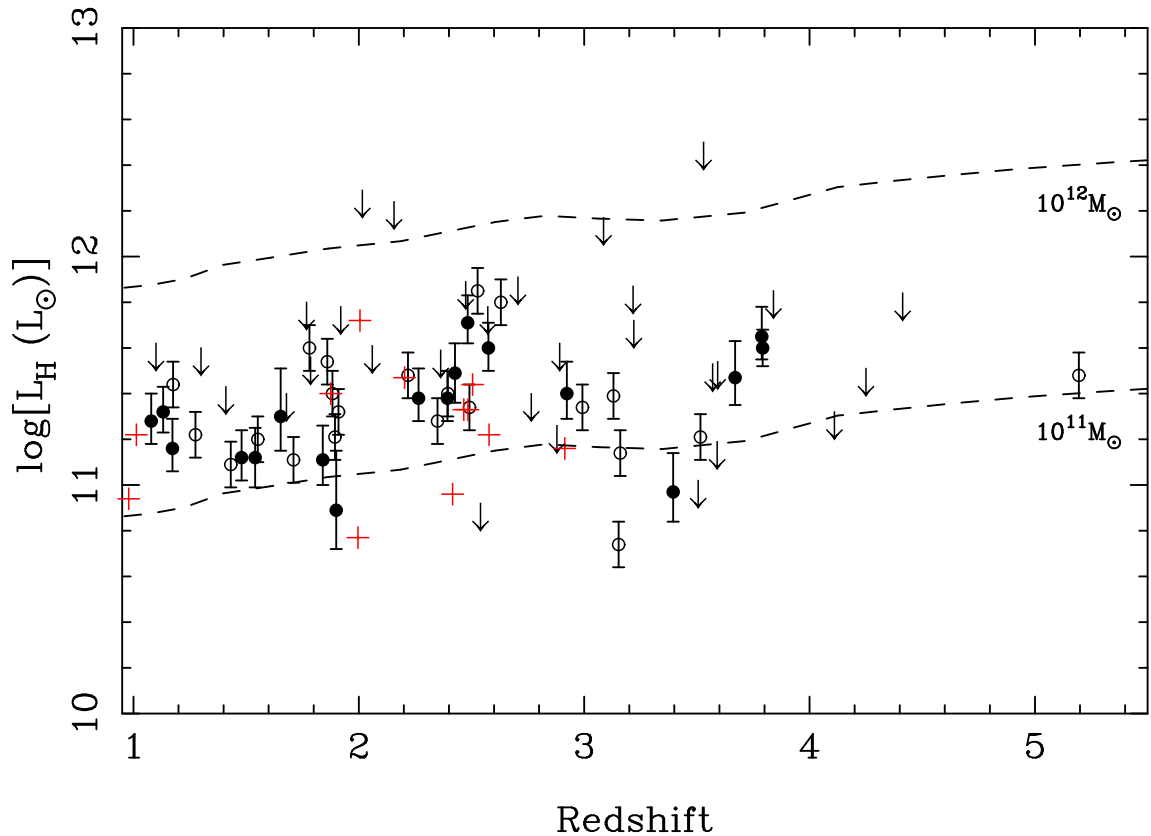}
\caption{Rest-frame $H-$\,band stellar luminosity versus redshift for the 
  {\it Spitzer} HzRG sample, derived from the best-fit models to the 
  multi-band photometry. Solid circles indicate those luminosities derived 
  from HzRGs with MIPS detections, whilst open circles are luminosities 
  derived from HzRGs without MIPS detections. Those radio galaxies where only 
  a maximum fit of the stellar SED was possible have upper limits, indicated 
  by arrows. The dashed lines represent the luminosities of elliptical 
  galaxies with $z_{\rm form}=10$ taken from the {\tt P\'EGASE.2} models and 
  normalized to $10^{11}M_\odot$ and $10^{12}M_\odot$. Crosses mark the 
  stellar luminosity of sub-millimetre galaxies from Borys et al. (2005), 
  re-derived in the same fashion as our HzRG sample.
\label{fig.logh}}
\end{figure}

\begin{figure}
\includegraphics[width=11cm,angle=270]{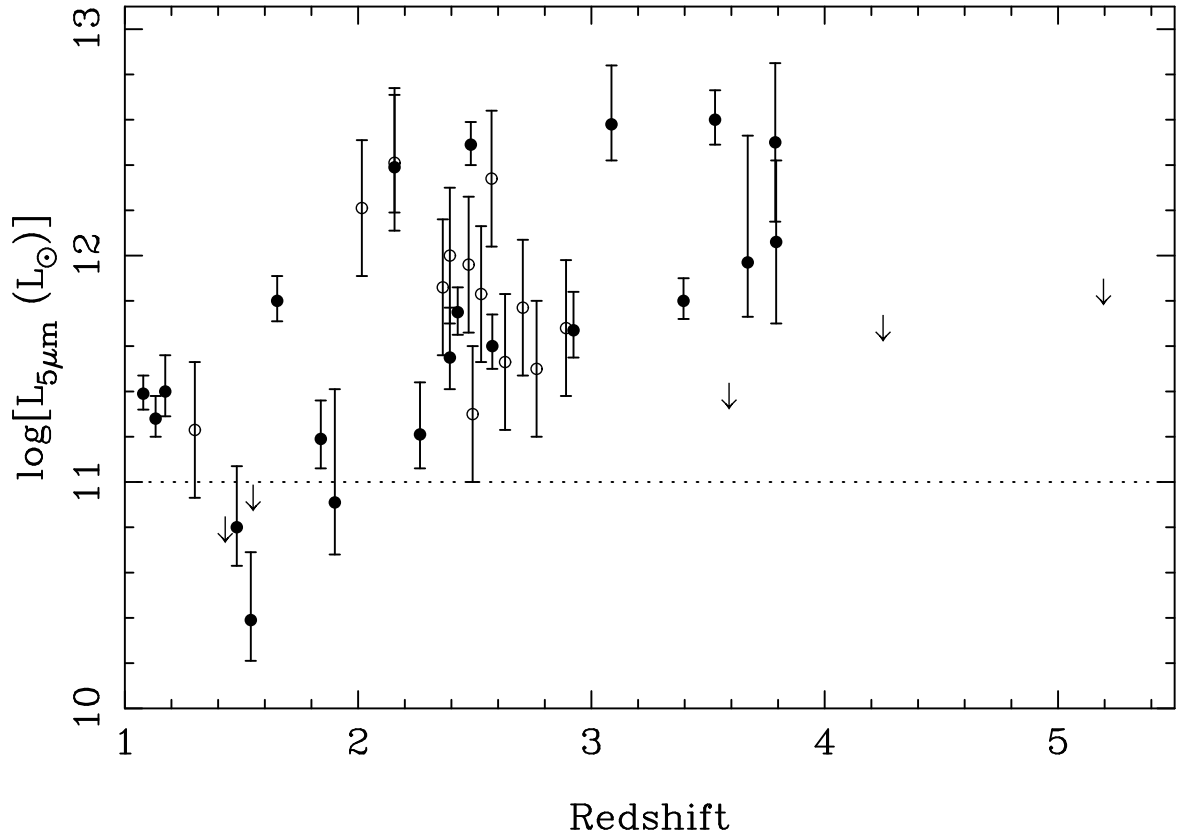}
\caption{Rest $5\,\um$ monochromatic luminosity versus redshift. 
  Filled symbols indicate sources observed with MIPS; open symbols
  indicate sources without MIPS observations but observed with IRS.
  Dotted line represents the divide between mid-IR ``strong'' and 
  mid-IR ``weak'' AGN, as defined by Ogle et al. (2006).
  \label{fig.z5um}}
\end{figure}

\begin{figure}
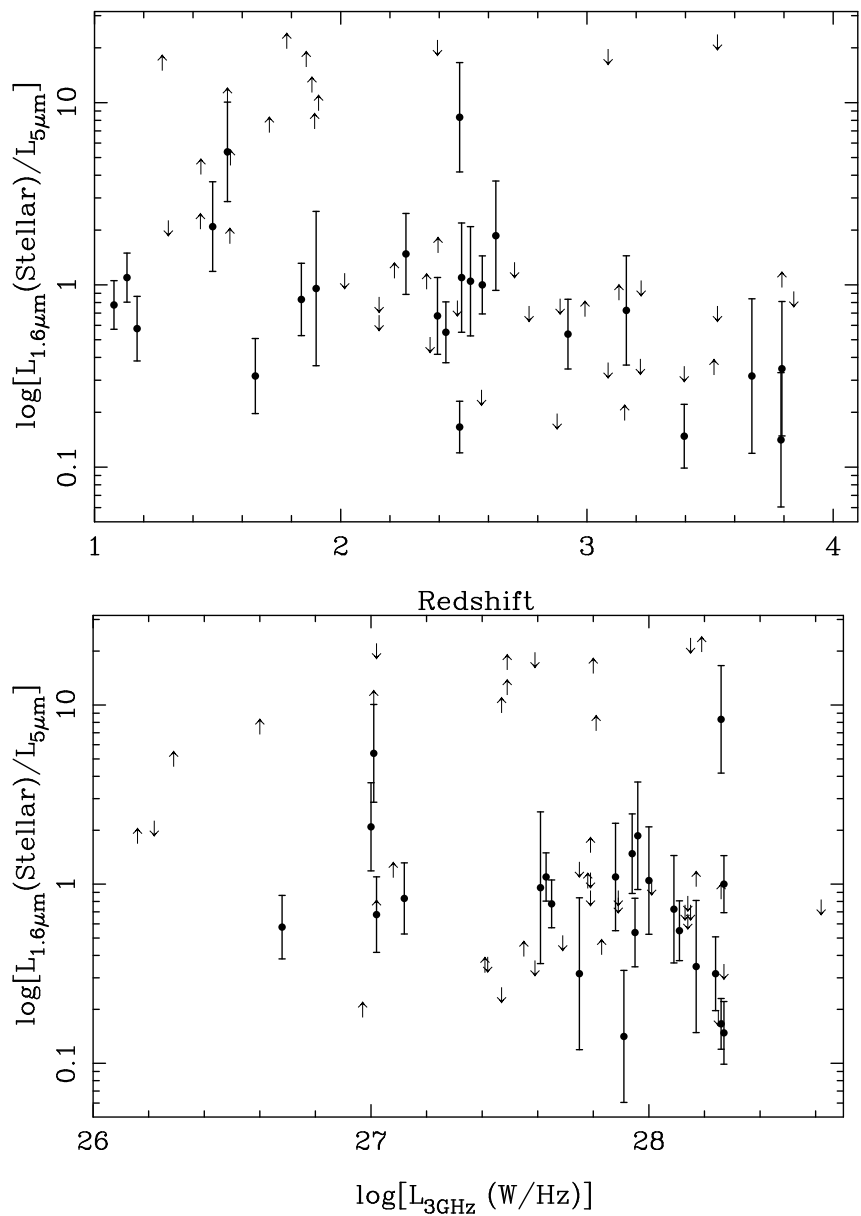

\includegraphics[width=8cm,angle=270]{f11a.ps}\\
\includegraphics[width=8cm,angle=270]{f11b.ps}
\caption{Ratio of rest-frame stellar near-IR ($H-$\,band) luminosity to 
  total mid-IR ($5\,\um$) luminosity against redshift (top) and $3\,$GHz 
  luminosity (bottom). This ratio appears to strongly correlate with both 
  redshift and radio luminosity, although both correlations are probably 
  due to the loose correlation of the $5\,\um$ luminosity
  with both redshift and radio luminosity.
  \label{fig.ratiof}}
\end{figure}

\begin{figure}
\includegraphics[width=13cm,angle=270]{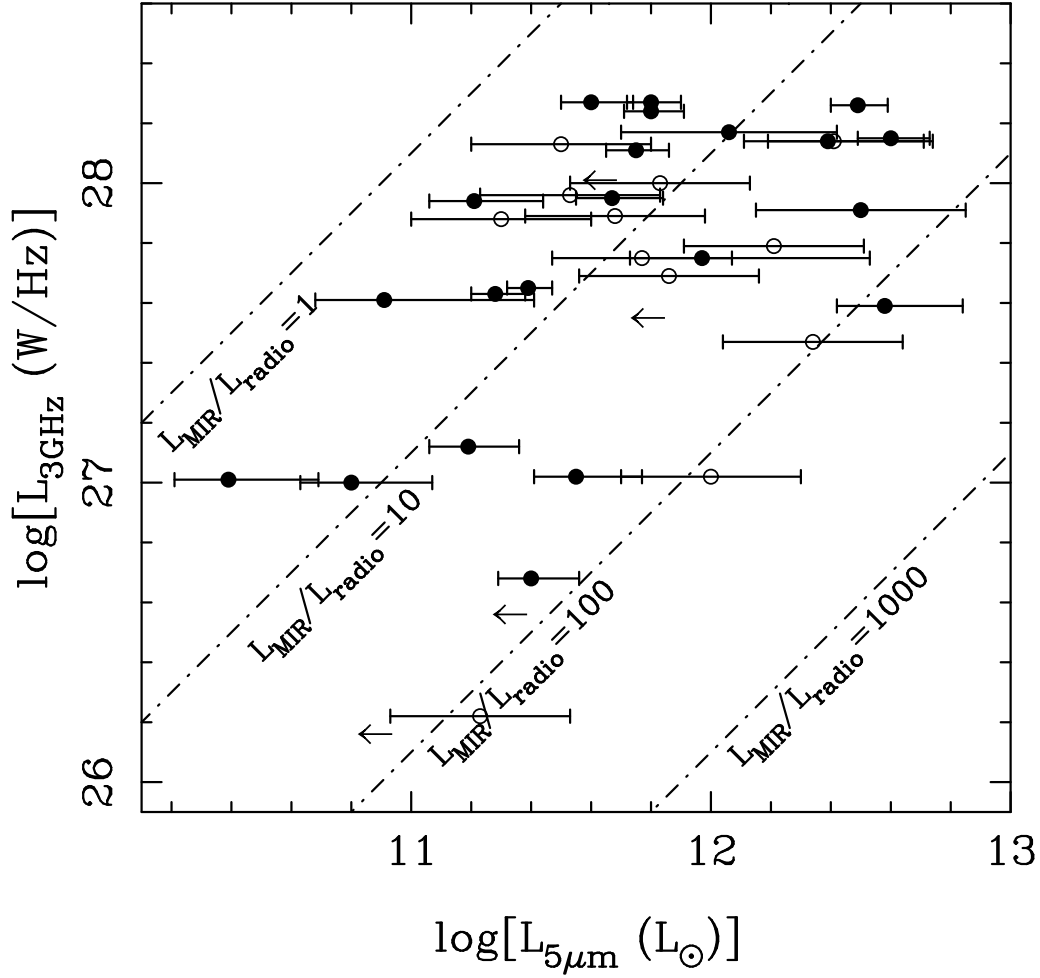}
\caption{Rest $5\,\um$ monochromatic luminosities versus rest 
  3\,GHz radio luminosities showing a slight correlation potentially
  due distance effects, or more likely demonstrating the both 
  mid-IR and radio luminosity trace AGN power and that the scatter is
  induced by the time delay of jets powering the lobes after a change 
  in AGN power.
  \label{fig.mir_rad}}
\end{figure}

\begin{figure}
\includegraphics[width=11cm,angle=270]{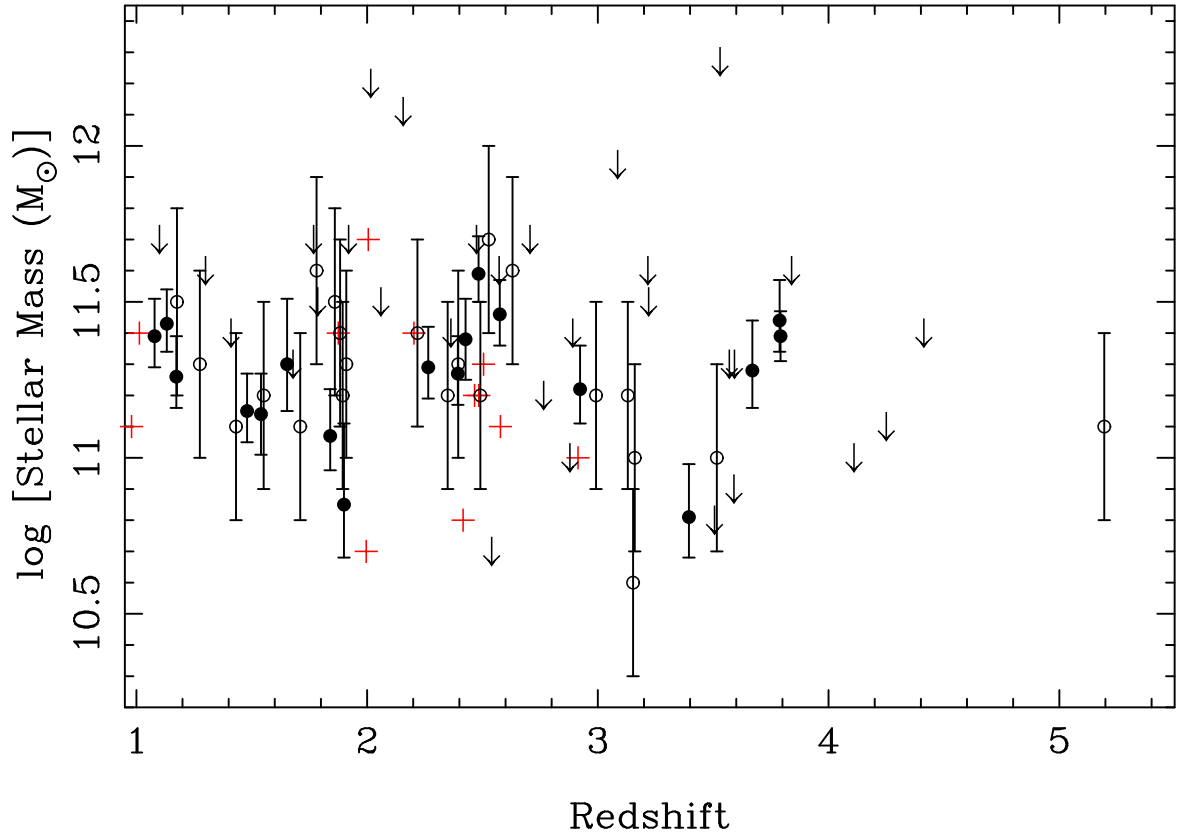}
\caption{Stellar mass in solar units plotted against redshift for the 69 
  {\it Spitzer} HzRGs,. Symbols are as in Fig.~\ref{fig.logh} 
\label{fig.logm}}
\end{figure}

\begin{figure}
\includegraphics[width=11cm,angle=270]{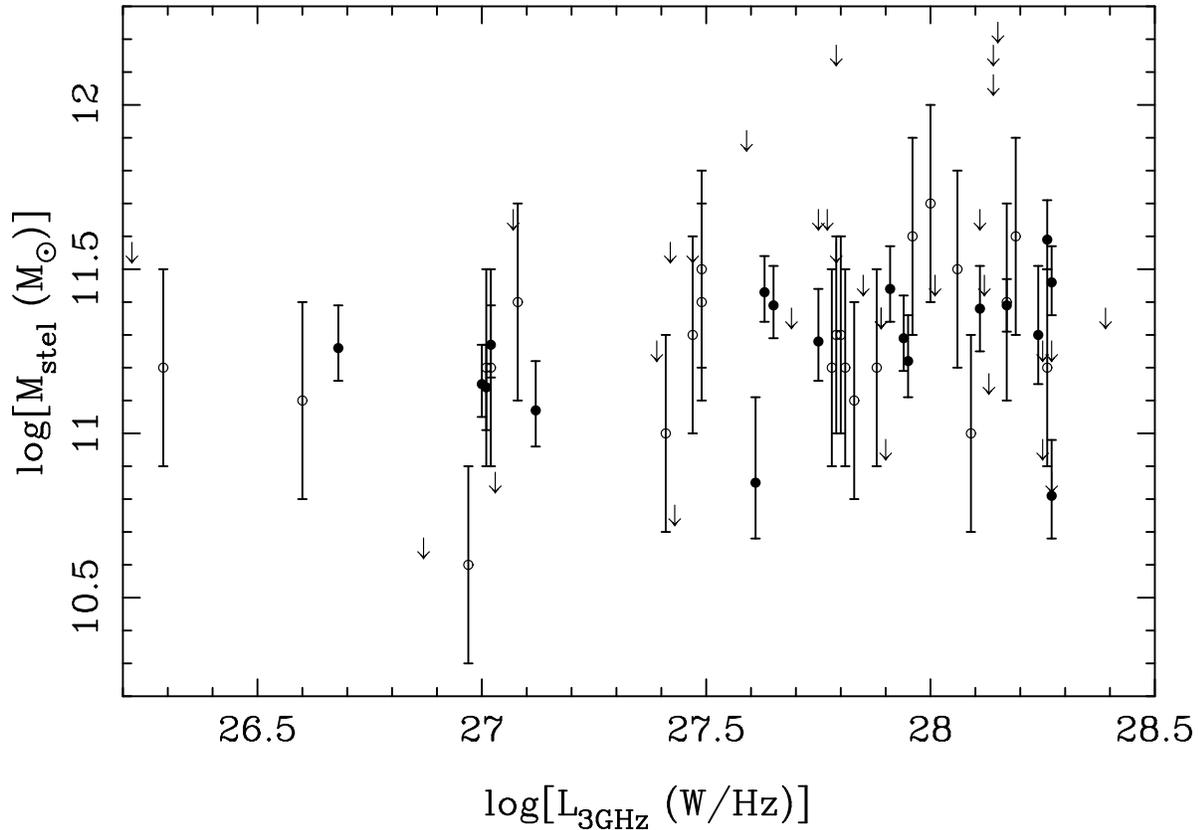}
\caption{Stellar mass against
  rest-frame 3\,GHz luminosity.  There is evidence of a slight trend,
  such that the most massive galaxies host more powerful radio sources,
  though, accounting for the large number of upper limits in the plot, a
  formal survival analysis shows only a 50\% chance of the two observables
  being correlated.  Deeper data and better coverage with the longer
  wavelength {\it Spitzer} cameras would show conclusively whether or not
  this correlation exists.
  \label{fig.radlogm}}
\end{figure}

Here we provide individual comments on selected HzRGs.  The SED fitting
is presented in Fig.~\ref{fig.mipsfit} (20 sources with MIPS detections)
and Fig.~\ref{fig.iracfit} (50 without MIPS detection), arranged in
order of increasing redshift.

\begin{itemize}

\item 3C65 ($z=1.176$): This HzRG was observed in only two bands due to an astrometric 
  error in NED which we have since had corrected.

\item 3C294 ($z=1.786$): While this radio galaxy is detected in all IRAC 
  bands, the photometry is severely affected by a bright star $8''$a to the 
  east by north-east. This bright star has allowed detailed adaptive optics
  imaging of this object \citep[e.g.][]{Quirrenbach:01, Steinbring:02}

\item B3~J2330+3927 ($z=3.086$): This HzRG is one of two sources consistent 
  with an AGN 
  power law dominating the entire optical to mid-IR emission. We therefore 
  can only provide an upper limit to the stellar mass. Although optical and 
  near-IR spectroscopy \citep{DeBreuck:03} show a classical type 2 AGN 
  spectrum, \citet{Pereztorres:05} found that the radio morphology is more 
  consistent with a type 1 AGN. The latter would be in agreement with the 
  high contribution from hot AGN light seen in the near- to mid-IR SED.

\item MRC~0316$-$257 ($z=3.131$): Due to an error at the time of proposal 
  submission, the 
  source observed with IRAC was MRC~0316$-$257B \citep{LeFevre:96}, a 
  spectroscopically confirmed $z=3.118$ AGN, $71''$ from MRC~0316$-$257. 
  We used the deeper Cycle~1 GO data (Program ID 3482, PI van Breugel), but 
  no IRS 16\,$\mu$m data is available for the $z>2$ radio galaxy.

\item TX~J1908+7220 ($z=3.53$): This HzRG is the second of two sources 
  consistent with an AGN power
  law dominating the entire optical to mid-IR emission. We therefore can
  only provide an upper limit to the stellar mass. 

\item TN~J1338$-$1942 ($z=4.11$): This HzRG is the fourth highest redshift 
  source in our sample \citep{DeBreuck:99}. Although the 3.6\,$\mu$m IRAC band 
  is likely to be contaminated by H$\alpha$, the remaining bands are 
  consistent with a stellar SED. The low 8.0\,$\mu$m flux ensures that any 
  hot dust contamination will be small.

\item 8C~1435+635 ($z=4.25$): This HzRG is the third highest redshift source 
  in our sample \citep{Lacy:94}. The stellar mass estimate is based on 
  the 4.5\,$\mu$m IRAC detection only, as the 3.6\,$\mu$m band is likely to 
  be contaminated by H$\alpha$ emission.

\item 6C~0140$+$326 ($z=4.41$): This HzRG is the second most distant galaxy 
  and is masked by a bright, foreground galaxy in IRAC channels one and two, 
  thus rendering source extraction at these wavelengths infeasible.

\item TN~J0924$-$2201 ($z=5.19$): This HzRG is the most distant radio galaxy 
  in our sample \citep{vanBreugel:99a}. It is detected only at 3.6 and 
  4.5\,$\mu$m. Deeper IRAC data has been obtained (Program ID 20749, PI 
  Zheng), but these data were not public yet at the time of writing (December
  2006).

\end{itemize}

\nocite{Songaila:94}
\nocite{Dickinson:03}
\nocite{Eales:96}
\nocite{Nesvadba:06}
\nocite{Devriendt:99}

%\input{bibinput}

% TABLE 1
\clearpage
\begin{deluxetable}{ccccccccc}
\tabletypesize{\tiny}
\tablecaption{{\it Spitzer} HzRG Sample and Exposure Times per Instrument}
\tablewidth{0pt}
\tablehead{
\colhead{} &
\colhead{RA} & 
\colhead{Dec.} & 
\colhead{} &
\colhead{$\log(L_{\rm 3GHz})$}  & 
\colhead{$K$-band$^a$} & 
\colhead{IRAC} &
\colhead{IRS} & 
\colhead{MIPS} \\
\colhead{HzRG} &
\colhead{(J2000)} & 
\colhead{(J2000)} & 
\colhead{Redshift} &
\colhead{(W Hz$^{-1}$)}  & 
\colhead{(Vega)} &
\colhead{(s)} &
\colhead{(s)} & 
\colhead{(s)}}
\startdata
    6C~0032+412 & 00:34:53.1 & $ 41$:31:31.50 & 3.670 & 27.75 & 19.10 &120 & 122 & 134, 420, 881 \\ 
   MRC~0037-258 & 00:39:56.4 & $-25$:34:31.01 & 1.100 & 27.07 & 17.10 &120 &  -  &  -  \\ 
    6C*0058+495 & 01:01:18.9 & $ 49$:50:12.29 & 1.173 & 26.68 & 17.60 &120 &  -  & 134, 420, 881 \\ 
   MRC~0114-211 & 01:16:51.4 & $-20$:52:06.71 & 1.410 & 28.39 & 18.50 &120 &  -  &  -  \\ 
  TN~J0121+1320 & 01:21:42.7 & $ 13$:20:58.00 & 3.516 & 27.41 & 18.70 &120 & 122 &  -  \\ 
    6C*0132+330 & 01:35:30.4 & $ 33$:17:00.82 & 1.710 & 26.60 & 18.80 &120 &  -  &  -  \\ 
    6C~0140+326 & 01:43:43.8 & $ 32$:53:49.31 & 4.413 & 27.89 & 19.90 &5000 & 122 & 267, 671, 2643 \\ 
   MRC~0152-209 & 01:54:55.8 & $-20$:40:26.30 & 1.920 & 27.77 & 17.90 &120 &  -  &  -  \\ 
   MRC~0156-252 & 01:58:33.6 & $-24$:59:31.10 & 2.016 & 27.79 & 16.10 &120 & 122 &  -  \\ 
  TN~J0205+2242 & 02:05:10.7 & $ 22$:42:50.40 & 3.506 & 27.43 & 18.70 &120 & 122 &  -  \\ 
   MRC~0211-256 & 02:13:30.5 & $-25$:25:21.00 & 1.300 & 26.22 &  - &120 & 122 &  -  \\ 
           3C65 & 02:23:43.2 & $ 40$:00:52.40 & 1.176 & 28.06 & 16.80 &120 &  -  &  -  \\ 
   MRC~0251-273 & 02:53:16.7 & $-27$:09:13.03 & 3.160 & 28.09 & 18.50 &120 & 122 &  -  \\ 
   MRC~0316-257 & 03:18:12.0 & $-25$:35:11.00 & 3.130 & 28.26 & 18.10 &120 & 122 &  -  \\ 
   MRC~0324-228 & 03:27:04.4 & $-22$:39:42.60 & 1.894 & 27.81 & 18.80 &120 &  -  &  -  \\ 
   MRC~0350-279 & 03:52:51.6 & $-27$:49:22.61 & 1.900 & 27.61 & 19.00 &120 &  -  & 134, 420, 881 \\ 
   MRC~0406-244 & 04:08:51.5 & $-24$:18:16.39 & 2.427 & 28.11 & 17.40 &120 & 122 & 134, 420, 881 \\ 
        4C60.07 & 05:12:54.8 & $ 60$:30:52.01 & 3.788 & 27.91 & 19.20 &120 & 122 & 134, 420, 881 \\ 
   PKS~0529-549 & 05:30:25.2 & $-54$:54:22.00 & 2.575 & 28.27 &  - &120 & 122 & 134, 420, 881 \\ 
  WN~J0617+5012 & 06:17:39.4 & $ 50$:12:55.40 & 3.153 & 26.97 & 19.70 &120 & 122 &  -  \\ 
        4C41.17 & 06:50:52.1 & $ 41$:30:31.00 & 3.792 & 28.17 & 19.10 &5000 & 122 & 267, 671, 2643 \\ 
  WN~J0747+3654 & 07:47:29.4 & $ 36$:54:38.09 & 2.992 & 27.02 & 20.00 &120 & 122 &  -  \\ 
   6CE0820+3642 & 08:23:48.1 & $ 36$:32:46.42 & 1.860 & 27.49 & 18.20 &120 &  -  &  -  \\ 
   USS~0828+193 & 08:30:53.4 & $ 19$:13:16.00 & 2.572 & 27.47 & 18.20 &120 & 122 &  -  \\ 
        5C7.269 & 08:28:38.8 & $ 25$:28:27.10 & 2.218 & 27.08 & 18.90 &120 & 122 &  -  \\ 
   6CE0901+3551 & 09:04:32.4 & $ 35$:39:03.23 & 1.910 & 27.47 & 18.10 &120 &  -  &  -  \\ 
   6CE0905+3955 & 09:08:16.9 & $ 39$:43:26.00 & 1.883 & 27.49 & 18.30 &120 &  -  &  -  \\ 
     B2~0902+34 & 09:05:30.1 & $ 34$:07:56.89 & 3.395 & 28.27 & 19.90 &1200 & 122 & 2557, 2696, 3556 \\ 
  TN~J0924-2201 & 09:24:19.9 & $-22$:01:41.00 & 5.195 & 27.83 & 19.70 &120 & 122 & 134, 420, 881 \\ 
    6C~0930+389 & 09:33:06.9 & $ 38$:41:50.14 & 2.395 & 27.79 & 19.50 &120 & 122 &  -  \\ 
   USS~0943-242 & 09:45:32.7 & $-24$:28:49.65 & 2.923 & 27.95 & 19.20 &120 & 122 & 134, 420, 881 \\ 
          3C239 & 10:11:45.4 & $ 46$:28:19.75 & 1.781 & 28.19 & 17.80 &120 &  -  &  -  \\ 
   MRC~1017-220 & 10:19:49.0 & $-22$:19:58.03 & 1.768 & 28.11 & 17.40 &120 &  -  &  -  \\ 
   MG~1019+0534 & 10:19:33.4 & $  5$:34:34.80 & 2.765 & 28.13 & 19.10 &120 & 122 &  -  \\ 
  WN~J1115+5016 & 11:15:06.9 & $ 50$:16:23.92 & 2.540 & 26.87 & 19.20 &120 & 122 &  -  \\ 
          3C257 & 11:23:09.2 & $  5$:30:19.47 & 2.474 & 28.62 & 17.80 &120 & 122 &  -  \\ 
  WN~J1123+3141 & 11:23:55.9 & $ 31$:41:26.14 & 3.217 & 27.42 & 17.40 &120 & 122 &  -  \\ 
   PKS~1138-262 & 11:40:48.6 & $-26$:29:08.50 & 2.156 & 28.14 & 16.10 &3000 & 122 & 9000, -, - \\ 
          3C266 & 11:45:43.4 & $ 49$:46:08.24 & 1.275 & 27.80 & 17.10 &120 &  -  &  -  \\ 
     6C~1232+39 & 12:35:04.8 & $ 39$:25:38.91 & 3.220 & 28.01 & 18.00 &120 & 122 &  -  \\ 
   USS~1243+036 & 12:45:38.4 & $  3$:23:20.70 & 3.570 & 28.25 & 19.20 &120 & 122 &  -  \\ 
  TN~J1338-1942 & 13:38:26.0 & $-19$:42:31.00 & 4.110 & 27.90 & 19.60 &5000 & 122 &  -  \\ 
        4C24.28 & 13:48:14.8 & $ 24$:15:52.00 & 2.879 & 28.25 &  -    &120 & 122 &  -  \\ 
        3C294.0 & 14:06:44.0 & $ 34$:11:25.00 & 1.786 & 28.12 & 17.90 &120 &  -  &  -  \\ 
   USS~1410-001 & 14:13:15.1 & $  0$:22:59.70 & 2.363 & 27.69 &  -    &120 & 122 &  -  \\ 
    8C~1435+635 & 14:36:37.1 & $ 63$:19:14.00 & 4.250 & 28.55 & 19.40 &120 & 122 & 134, 420, 881 \\ 
   USS~1558-003 & 16:01:17.3 & $  0$:28:48.00 & 2.527 & 28.00 &   -   &120 & 122 &  -  \\ 
   USS~1707+105 & 17:10:06.5 & $ 10$:31:06.00 & 2.349 & 27.78 &   -   &120 & 122 &  -  \\ 
    LBDS~53w002 & 17:14:14.7 & $ 50$:15:29.70 & 2.393 & 27.02 & 18.90 &3300 & 122 & 134, 420, 881 \\ 
    LBDS~53w091 & 17:22:32.7 & $ 50$:06:01.94 & 1.552 & 26.29 & 18.70 &900 &  -  & 633, 1311, 2643 \\ 
        3C356.0 & 17:24:19.0 & $ 50$:57:40.30 & 1.079 & 27.65 & 16.80 &120 &  -  & 134, 420, 881 \\ 
   7C~1751+6809 & 17:50:49.9 & $ 68$:08:25.93 & 1.540 & 27.01 & 18.20 &120 &  -  & 134, 420, 881 \\ 
   7C~1756+6520 & 17:57:05.4 & $ 65$:19:53.11 & 1.480 & 27.00 & 18.90 &120 &  -  & 134, 420, 881 \\ 
        3C368.0 & 18:05:06.3 & $ 11$:01:33.00 & 1.132 & 27.63 & 17.20 &120 &  -  & 134, 420, 881 \\ 
   7C~1805+6332 & 18:05:56.9 & $ 63$:33:13.14 & 1.840 & 27.12 & 18.80 &120 &  -  & 134, 420, 881 \\ 
        4C40.36 & 18:10:55.7 & $ 40$:45:24.01 & 2.265 & 27.94 & 17.80 &120 & 122 & 134, 420, 881 \\ 
  TX~J1908+7220 & 19:08:23.7 & $ 72$:20:11.82 & 3.530 & 28.15 & 16.50 &120 & 122 & 134, 420, 881 \\ 
  WN~J1911+6342 & 19:11:49.6 & $ 63$:42:09.60 & 3.590 & 27.03 & 19.90 &120 & 122 & 134, 420, 881 \\ 
  TN~J2007-1316 & 20:07:53.3 & $-13$:16:43.62 & 3.840 & 27.79 & 18.80 &120 & 122 &  -  \\ 
   MRC~2025-218 & 20:27:59.5 & $-21$:40:56.90 & 2.630 & 27.96 & 18.50 &120 & 122 &  -  \\ 
   MRC~2048-272 & 20:51:03.6 & $-27$:03:02.53 & 2.060 & 27.85 & 18.30 &120 & 122 &  -  \\ 
   MRC~2104-242 & 21:06:58.1 & $-24$:05:11.00 & 2.491 & 27.88 &   -   &120 & 122 &  -  \\ 
        4C23.56 & 21:07:14.8 & $ 23$:31:45.00 & 2.483 & 28.26 & 19.73 &120 & 122 & 134, 420, 881 \\ 
   MG~2144+1928 & 21:44:07.5 & $ 19$:29:14.60 & 3.592 & 28.27 & 19.10 &120 & 122 &  -  \\ 
   USS~2202+128 & 22:05:14.1 & $ 13$:05:33.50 & 2.706 & 27.75 & 18.40 &120 & 122 &  -  \\ 
   MRC~2224-273 & 22:27:43.3 & $-27$:05:01.71 & 1.679 & 27.39 & 18.50 &120 &  -  &  -  \\ 
  B3~J2330+3927 & 23:30:24.9 & $ 39$:27:12.02 & 3.086 & 27.59 & 18.80 &120 & 122 & 134, 420, 881 \\ 
        4C28.58 & 23:51:59.2 & $ 29$:10:28.99 & 2.891 & 27.89 & 18.70 &120 & 122 &  -  \\ 
          3C470 & 23:58:35.3 & $ 44$:04:38.87 & 1.653 & 28.24 & 18.50 &120 &  -  & 134, 420, 881 \\ 
\enddata
\tablenotetext{a}{$K$-band magnitude within projected 64\,kpc radius}
\normalsize
\label{table.shzrg}
\end{deluxetable}
%\addtocounter{table}{-1}

%\begin{deluxetable}{ccccccccc}
%\tiny
%\tablecaption{- (continued) {\it Spitzer} HzRG Sample and Exposure Times per Instrument}
%\tablewidth{0pt}
%\tablehead{
%\colhead{} &
%\colhead{RA} & 
%\colhead{Dec.} & 
%\colhead{} &
%\colhead{$\log(L_{\rm 3GHz}$)}  & 
%\colhead{$K$-band$^a$} & 
%\colhead{IRAC} &
%\colhead{IRS} & 
%\colhead{MIPS} \\
%\colhead{HzRG} &
%\colhead{(J2000)} & 
%\colhead{(J2000)} & 
%\colhead{Redshift} &
%\colhead{(W Hz$^{-1}$)}  & 
%\colhead{(Vega)} &
%\colhead{(s)} &
%\colhead{(s)} & 
%\colhead{(s)}}
%\startdata
%\input{tab1b}
%\enddata
%\tablenotetext{a}{$K$-band magnitude within projected 64\,kpc}
%\normalsize
%%\label{table.shzrg}
%\end{deluxetable}

% TABLE 2
\clearpage
\begin{deluxetable}{ccccccc}
\footnotesize
\tablecaption{{\it Spitzer} Photometric Parameters} 
\tablewidth{0pt}
\tablehead{
\colhead{} & 
\colhead{Wavelength} & 
\colhead{Pixel Scale} &
\colhead{Conv.} & 
\colhead{Aperture$^{a}$} & 
\colhead{Annulus$^{b}$} & 
\colhead{Aperture} \\ 
\colhead{Filter} & 
\colhead{($\um$)} & 
\colhead{(\arcsec/pix)} & 
\colhead{to $\mu$Jy} & 
\colhead{(\arcsec)} &
\colhead{(\arcsec)} & 
\colhead{Correction}}
\startdata
IRAC1      & 3.6 & 0.61 & 23.5       & 3.5 & 9.76   & 1.48     \\ 
           &     &      &            & 7.0 & 9.76   & 1.13   \\ 
IRAC2      & 4.5 & 0.61 & 23.5       & 3.5 & 9.76   & 1.60   \\ 
           &     &      &            & 7.0 & 9.76   & 1.168   \\ 
IRAC3      & 5.8 & 0.61 & 23.5       & 3.5 & 9.76   & 2.76     \\ 
           &     &      &            & 7.0 & 9.76   & 1.155     \\ 
IRAC4      & 8.0 & 0.61 & 23.5       & 3.5 & 9.76   & 2.13     \\ 
           &     &      &            & 7.0 & 9.76   & 1.316     \\ 
IRS-blue   & 16  & 1.2  & 0.591$^{c}$ & 6  & 6-10  & -      \\ 
MIPS1      & 24  & 1.25 & 36.71       & 13 & 20-32 & 1.167  \\
MIPS2      & 70  & 4.95 & 576         & 35 & 39-65 & 1.211$^{d}$  \\
MIPS3      & 160 & 8    & 1500        & 50 & 75-125 & 1.445 \\
\enddata
\tablenotetext{a}{Diameter in which the photometry is measured.}
\tablenotetext{b}{Inner and outer diameter of annulus used to 
  determine the background sky level, except for the IRAC bands 
  where we provide the size of the background mesh used by
  {\tt SExtractor}.}
\tablenotetext{c}{Including aperture correction.}
\tablenotetext{d}{For a power-law SED,  $f_\nu\propto\nu^{-2}$.}
\label{table.irsmips}
\end{deluxetable}

% TABLE 3
\clearpage
\begin{deluxetable}{ccccccccc}
\tabletypesize{\tiny}
\tablecaption{{\it Spitzer} Photometry for HzRGs ($\uJy$)}
\tablewidth{0pt}
\tablehead{
\colhead{HzRG} &
\colhead{$f_{3.6\um}$} &
\colhead{$f_{4.5\um}$} &
\colhead{$f_{5.8\um}$} &
\colhead{$f_{8.0\um}$} &
\colhead{$f_{16\um}$} &
\colhead{$f_{24\um}$} &
\colhead{$f_{70\um}$} &
\colhead{$f_{160\um}$}}
\startdata
   6C~0032+412 &   15.4 $\pm$    2.1 &   33.8 $\pm$    3.7 &   57.6 $\pm$    8.9 &   98.3 $\pm$   10.0 &  205.0 $\pm$   34.9 &  457.0 $\pm$   59.7 & $<  3950$ & $<  96000$ \\ 
  MR~C0037-258 &  221.0 $\pm$   22.0 &  248.0 $\pm$   25.0 &  286.0 $\pm$   29.0 &  518.0 $\pm$   52.0 &    $-$ &    $-$ &    $-$ &     $-$ \\ 
   6C*0058+495 &   82.0 $\pm$    8.3 &   86.7 $\pm$    8.8 &   93.2 $\pm$    9.5 &  309.0 $\pm$   31.0 &    $-$ & 1450.0 $\pm$   87.3 &  18900 $\pm$   1962 & $<  94600$ \\ 
  MRC~0114-211 &   87.3 $\pm$    8.9 &  117.0 $\pm$   12.0 &  157.0 $\pm$   16.0 &  398.0 $\pm$   40.0 &    $-$ &    $-$ &    $-$ &     $-$ \\ 
 TN~J0121+1320 &   10.5 $\pm$    1.8 &   14.4 $\pm$    2.1 & $<  32.5$ & $<  40.3$ & $< 136.0$ &    $-$ &    $-$ &     $-$ \\ 
   6C*0132+330 &   31.6 $\pm$    3.6 &   41.3 $\pm$    4.5 &   51.9 $\pm$   12.4 &  109.0 $\pm$   11.0 &    $-$ &    $-$ &    $-$ &     $-$ \\ 
   6C~0140+326 & $< 623.0$ & $< 450.0$ &   34.9 $\pm$    2.7 &   36.3 $\pm$    3.4 &  194.0 $\pm$   40.0 & 2190.0 $\pm$  110.5 & $<  2580$ & $<  20700$ \\ 
  MRC~0152-209 &  108.0 $\pm$   11.0 &  165.0 $\pm$   17.0 &  215.0 $\pm$   22.0 &  415.0 $\pm$   42.0 &    $-$ &    $-$ &    $-$ &     $-$ \\ 
  MRC~0156-252 &  291.0 $\pm$   29.0 &  405.0 $\pm$   41.0 &  717.0 $\pm$   72.0 & 1125.0 $\pm$  113.0 & 1430.0 $\pm$  116.6 &    $-$ &    $-$ &     $-$ \\ 
 TN~J0205+2242 &    7.5 $\pm$    1.7 &    4.7 $\pm$    2.3 & $<  29.3$ & $<  37.4$ & $< 122.0$ &    $-$ &    $-$ &     $-$ \\ 
  MRC~0211-256 &  166.0 $\pm$   17.0 &  197.0 $\pm$   20.0 &  222.0 $\pm$   22.0 &  278.0 $\pm$   28.0 &  330.0 $\pm$   42.6 &    $-$ &    $-$ &     $-$ \\ 
          3C65 &  161.0 $\pm$   32.0 &    $-$ &  320.0 $\pm$   66.0 &    $-$ &    $-$ &    $-$ &    $-$ &     $-$ \\ 
  MRC~0251-273 &   11.5 $\pm$    1.6 &   12.7 $\pm$    1.8 & $<  29.1$ & $<  28.8$ &   97.2 $\pm$   34.2 &    $-$ &    $-$ &     $-$ \\ 
  MRC~0316-257 &   19.3 $\pm$    2.1 &   20.1 $\pm$    2.1 &   19.5 $\pm$    2.4 &   38.1 $\pm$    4.1 & $<  91.5$ &    $-$ &    $-$ &     $-$ \\ 
  MRC~0324-228 &   39.4 $\pm$    4.2 &   39.7 $\pm$    4.3 &   61.1 $\pm$    8.6 &   89.9 $\pm$    9.9 &    $-$ &    $-$ &    $-$ &     $-$ \\ 
  MRC~0350-279 &   23.6 $\pm$    2.6 &   40.6 $\pm$    4.2 &   82.2 $\pm$   27.6 &   79.3 $\pm$   26.5 &    $-$ &  321.0 $\pm$   60.7 & $<  1040$ & $<  88800$ \\ 
  MRC~0406-244 &   40.4 $\pm$    4.3 &   43.3 $\pm$    4.6 & $<  51.6$ &   63.5 $\pm$   14.5 &  242.0 $\pm$   34.6 & 1420.0 $\pm$   83.7 &  24700 $\pm$   2306 & $<  47700$ \\ 
       4C60.07 &   27.3 $\pm$    3.1 &   33.6 $\pm$    3.7 &   35.1 $\pm$    8.4 &   37.2 $\pm$    9.5 &  175.0 $\pm$   33.0 & 1340.0 $\pm$   83.7 & $<  3750$ & $<  64600$ \\ 
  PKS~0529-549 &   46.6 $\pm$    4.9 &   52.9 $\pm$    5.5 &   62.7 $\pm$    8.6 &   72.2 $\pm$    9.1 &  134.0 $\pm$   33.7 &  942.0 $\pm$   71.0 & $<  4110$ & $<  74100$ \\ 
 WN~J0617+5012 &    3.6 $\pm$    1.0 &    5.5 $\pm$    1.2 & $<  48.3$ & $<  54.5$ & $<  93.4$ &    $-$ &    $-$ &     $-$ \\ 
       4C41.17 &   23.4 $\pm$    2.4 &   27.5 $\pm$    2.8 &   35.6 $\pm$    3.7 &   36.5 $\pm$    3.5 & $<  99.3$ &  370.0 $\pm$   43.3 & $<  3210$ & $<  26300$ \\ 
 WN~J0747+3654 &   19.1 $\pm$    2.4 &   25.3 $\pm$    3.0 & $<  29.9$ &   44.9 $\pm$   11.5 & $< 111.0$ &    $-$ &    $-$ &     $-$ \\ 
  6CE0820+3642 &   79.2 $\pm$    8.1 &   81.9 $\pm$    8.4 &   82.0 $\pm$    8.4 &   68.0 $\pm$    7.0 &    $-$ &    $-$ &    $-$ &     $-$ \\ 
  USS~0828+193 &   61.7 $\pm$    6.9 &  133.0 $\pm$   13.0 &  201.0 $\pm$   21.0 &  687.0 $\pm$   74.0 & 1270.0 $\pm$  112.8 &    $-$ &    $-$ &     $-$ \\ 
       5C7.269 &   41.0 $\pm$    4.5 &   49.5 $\pm$    5.3 &   57.8 $\pm$    6.1 & $<  40.1$ & $< 186.0$ &    $-$ &    $-$ &     $-$ \\ 
  6CE0901+3551 &   37.2 $\pm$    4.1 &   46.5 $\pm$    5.0 &   52.8 $\pm$   10.7 &   69.8 $\pm$   12.4 &    $-$ &    $-$ &    $-$ &     $-$ \\ 
  6CE0905+3955 &   51.8 $\pm$    5.4 &   60.1 $\pm$    6.2 &   96.8 $\pm$    9.8 &  146.0 $\pm$   14.0 &    $-$ &    $-$ &    $-$ &     $-$ \\ 
    B2~0902+34 &    6.4 $\pm$    0.8 &    9.9 $\pm$    1.1 &   11.0 $\pm$    2.5 &   41.3 $\pm$    2.3 &  144.0 $\pm$   39.6 &  323.0 $\pm$   18.8 &    $-$ & $<  18500$ \\ 
 TN~J0924-2201 &   11.3 $\pm$    1.8 &   11.4 $\pm$    1.9 & $<  30.6$ & $<  33.0$ & $< 108.0$ & $< 160.0$ & $<  3330$ & $<  52900$ \\ 
   6C~0930+389 &   30.7 $\pm$    3.4 &   32.2 $\pm$    3.6 &   37.3 $\pm$    9.2 & $<  32.6$ & $<  97.2$ &    $-$ &    $-$ &     $-$ \\ 
  USS~0943-242 &   21.5 $\pm$    2.6 &   28.4 $\pm$    3.2 & $<  30.9$ &   25.8 $\pm$   11.7 &  109.0 $\pm$   34.1 &  493.0 $\pm$   57.5 & $<  3390$ & $<  50900$ \\ 
         3C239 &   96.4 $\pm$    9.8 &  111.0 $\pm$   11.0 &  130.0 $\pm$   12.0 &  142.0 $\pm$   14.0 &    $-$ &    $-$ &    $-$ &     $-$ \\ 
  MRC~1017-220 &  119.0 $\pm$   12.0 &  179.0 $\pm$   18.0 &  273.0 $\pm$   27.0 &  360.0 $\pm$   36.0 &    $-$ &    $-$ &    $-$ &     $-$ \\ 
  MG~1019+0534 &   25.6 $\pm$    2.9 &   19.5 $\pm$    3.8 & $<  35.4$ & $<  42.9$ &  161.0 $\pm$   52.5 &    $-$ &    $-$ &     $-$ \\ 
 WN~J1115+5016 &    7.8 $\pm$    2.1 &    9.5 $\pm$    3.0 & $<  54.8$ & $<  61.1$ & $<  88.0$ &    $-$ &    $-$ &     $-$ \\ 
         3C257 &   85.0 $\pm$    8.7 &  111.0 $\pm$   11.0 &  194.0 $\pm$   19.0 &  322.0 $\pm$   33.0 &  569.0 $\pm$   74.6 &    $-$ &    $-$ &     $-$ \\ 
 WN~J1123+3141 &   48.2 $\pm$    5.0 &   74.4 $\pm$    7.6 &   92.7 $\pm$    9.4 &  182.6 $\pm$   18.4 &  742.0 $\pm$   70.4 &    $-$ &    $-$ &     $-$ \\ 
  PKS~1138-262 &  318.0 $\pm$   32.0 &  497.0 $\pm$   50.0 &  887.0 $\pm$   89.0 & 1500.0 $\pm$  150.0 & 2000.0 $\pm$  160.3 & 3890.0 $\pm$  176.2 &    $-$ &     $-$ \\ 
         3C266 &   67.9 $\pm$    7.0 &   73.1 $\pm$    7.5 &   45.1 $\pm$    4.7 &  102.6 $\pm$   10.4 &    $-$ &    $-$ &    $-$ &     $-$ \\ 
    6C~1232+39 &   33.3 $\pm$    3.6 &   41.8 $\pm$    4.4 &   52.0 $\pm$    9.2 &   75.2 $\pm$   10.8 &  208.0 $\pm$   35.9 &    $-$ &    $-$ &     $-$ \\ 
  USS~1243+036 &   22.0 $\pm$    2.5 &   21.5 $\pm$    3.0 & $<  48.0$ & $<  62.9$ & $< 163.0$ &    $-$ &    $-$ &     $-$ \\ 
 TN~J1338-1942 &   17.8 $\pm$    1.9 &   10.7 $\pm$    1.2 &   14.3 $\pm$    2.8 &    9.9 $\pm$    3.4 & $< 119.0$ &    $-$ &    $-$ &     $-$ \\ 
       4C24.28 &   16.5 $\pm$    2.2 &   27.3 $\pm$    3.1 &   43.3 $\pm$    8.3 &  102.0 $\pm$   10.0 &  446.0 $\pm$   50.8 &    $-$ &    $-$ &     $-$ \\ 
       3C294.0 & $<  93.0$ & $< 103.0$ &   68.0 $\pm$   16.8 &   66.6 $\pm$   20.6 &    $-$ &    $-$ &    $-$ &     $-$ \\ 
  USS~1410-001 &   50.6 $\pm$    5.3 &   79.0 $\pm$    8.1 &  166.0 $\pm$   17.0 &  240.0 $\pm$   24.0 &  481.0 $\pm$   58.6 &    $-$ &    $-$ &     $-$ \\ 
    8C1435+635 &   18.6 $\pm$    2.1 &   14.7 $\pm$    2.9 & $<  45.5$ & $<  50.3$ & $<  86.7$ & $< 165.0$ & $<  4130$ & $<  33900$ \\ 
  USS~1558-003 &   78.8 $\pm$    8.1 &  101.0 $\pm$   10.3 &  105.0 $\pm$   10.5 &  233.0 $\pm$   23.4 &  398.0 $\pm$   52.2 &    $-$ &    $-$ &     $-$ \\ 
  USS~1707+105 &   22.1 $\pm$    2.7 &   30.1 $\pm$    3.4 &   22.6 $\pm$    9.1 & $<  33.3$ & $< 123.0$ &    $-$ &    $-$ &     $-$ \\ 
   LBDS~53w002 &   32.0 $\pm$    3.3 &   44.0 $\pm$    4.5 &   49.9 $\pm$    5.2 &  103.0 $\pm$   11.0 &  177.0 $\pm$   33.9 &  648.0 $\pm$   61.4 & $<  4300$ & $<  65100$ \\ 
   LBDS~53w091 &   43.3 $\pm$    6.3 &   51.4 $\pm$    5.3 &   23.9 $\pm$    6.1 &   26.6 $\pm$    6.4 &    $-$ & $<  45.9$ & $<  1610$ & $<  23900$ \\ 
       3C356.0 &  108.0 $\pm$   11.0 &  110.0 $\pm$   11.0 &  122.0 $\pm$   14.0 &  434.0 $\pm$   47.0 &    $-$ & 4060.0 $\pm$  192.3 & $<  4400$ & $<  70200$ \\ 
  7C~1751+6809 &   46.6 $\pm$    4.9 &   50.8 $\pm$    5.3 & $<  40.9$ &   36.5 $\pm$   16.0 &    $-$ &  351.0 $\pm$   50.5 & $<  3600$ & $<  51600$ \\ 
  7C~1756+6520 &   39.6 $\pm$    4.2 &   46.9 $\pm$    5.0 &   34.7 $\pm$    7.7 &   42.6 $\pm$    8.6 &    $-$ &  444.0 $\pm$   51.1 & $<  5900$ & $< 151000$ \\ 
       3C368.0 &  126.0 $\pm$   13.0 &  112.0 $\pm$   11.0 &  112.0 $\pm$   11.0 &  210.0 $\pm$   21.0 &    $-$ & 3250.0 $\pm$  166.7 &  28800 $\pm$   2710 & $<  39000$ \\ 
  7C~1805+6332 &   28.4 $\pm$    3.6 &   42.1 $\pm$    4.4 &   51.4 $\pm$    5.4 &   95.6 $\pm$   17.1 &    $-$ &  648.0 $\pm$   56.2 & $<  4330$ & $< 111000$ \\ 
       4C40.36 &   36.5 $\pm$    3.9 &   41.3 $\pm$    4.3 &   45.4 $\pm$   12.9 &   26.3 $\pm$   10.3 &  115.0 $\pm$   29.4 &  520.0 $\pm$   55.2 & $<  4750$ & $<  64500$ \\ 
 TX~J1908+7220 &  200.0 $\pm$   20.0 &  229.0 $\pm$   23.0 &  241.0 $\pm$   25.0 &  480.0 $\pm$   48.0 &  841.0 $\pm$   70.6 & 1910.0 $\pm$   98.9 &  16200 $\pm$   1905 & $<  63300$ \\ 
 WN~J1911+6342 &    9.5 $\pm$    1.8 &    8.7 $\pm$    1.8 & $<  26.4$ & $<  26.0$ & $<  75.3$ & $< 142.0$ & $<  3710$ & $<  57100$ \\ 
 TN~J2007-1316 &   56.2 $\pm$    5.9 &   54.4 $\pm$    5.8 &   41.1 $\pm$   10.7 &  121.3 $\pm$   12.3 &  244.0 $\pm$   53.5 &    $-$ &    $-$ &     $-$ \\ 
  MRC~2025-218 &   68.4 $\pm$    7.1 &   77.1 $\pm$    8.0 &   86.8 $\pm$   10.9 &  126.8 $\pm$   12.9 &  190.0 $\pm$   50.2 &    $-$ &    $-$ &     $-$ \\ 
  MRC~2048-272 &   59.5 $\pm$    6.2 &   72.6 $\pm$    7.5 &   78.3 $\pm$   10.4 &   38.4 $\pm$   13.1 & $< 141.0$ &    $-$ &    $-$ &     $-$ \\ 
  MRC~2104-242 &   28.1 $\pm$    3.3 &   29.7 $\pm$    3.5 &   32.8 $\pm$   10.0 & $<  36.3$ &  121.0 $\pm$   47.0 &    $-$ &    $-$ &     $-$ \\ 
       4C23.56 &   61.1 $\pm$    6.4 &   86.2 $\pm$    8.8 &  126.9 $\pm$   12.8 &  423.7 $\pm$   42.5 & 1610.0 $\pm$  129.8 & 4390.0 $\pm$  203.8 &  30300 $\pm$   2958 & $<  70500$ \\ 
  MG~2144+1928 &   22.1 $\pm$    2.7 &   18.3 $\pm$    2.4 & $<  25.5$ & $<  30.3$ & $<  93.9$ &    $-$ &    $-$ &     $-$ \\ 
  USS~2202+128 &   60.4 $\pm$    6.3 &   95.8 $\pm$    9.8 &  120.1 $\pm$   12.1 &  106.6 $\pm$   10.8 &  307.0 $\pm$   43.2 &    $-$ &    $-$ &     $-$ \\ 
  MRC~2224-273 &   61.6 $\pm$    6.4 &   86.1 $\pm$    8.8 &   98.4 $\pm$   10.0 &  203.9 $\pm$   20.5 &    $-$ &    $-$ &    $-$ &     $-$ \\ 
 B3~J2330+3927 &   99.6 $\pm$   10.1 &  143.0 $\pm$   14.0 &  160.0 $\pm$   16.0 &  474.0 $\pm$   47.0 & 1040.0 $\pm$   88.0 & 2210.0 $\pm$  113.2 & $<  4670$ & $<  64300$ \\ 
       4C28.58 &   31.6 $\pm$    3.5 &   36.0 $\pm$    3.9 & $<  41.7$ &   40.9 $\pm$    4.4 &  228.0 $\pm$   31.5 &    $-$ &    $-$ &     $-$ \\ 
         3C470 &   49.5 $\pm$   10.4 &   75.2 $\pm$   11.8 &   70.9 $\pm$   10.4 &  266.0 $\pm$   30.0 &    $-$ & 2650.0 $\pm$  133.5 & $<  5570$ & $< 102000$ \\ 
\enddata
\label{table.spitzer}
\end{deluxetable}
%\addtocounter{table}{-1}

%\clearpage
%\begin{deluxetable}{ccccccccc}
%\tiny
%\tablecaption{(continued) {\it Spitzer} Photometry for HzRGs ($\uJy$)}
%\tablewidth{0pt}
%\tablehead{
%\colhead{HzRG} &
%\colhead{$f_{3.6\um}$} &
%\colhead{$f_{4.5\um}$} &
%\colhead{$f_{5.8\um}$} &
%\colhead{$f_{8.0\um}$} &
%\colhead{$f_{16\um}$} &
%\colhead{$f_{24\um}$} &
%\colhead{$f_{70\um}$} &
%\colhead{$f_{160\um}$}}
%\startdata
%\input{tab3b}
%\enddata
%\end{deluxetable}

% TABLE 4
\clearpage
\begin{deluxetable}{ccccccc}
\tablecaption{Results of SED fitting for HzRGs with MIPS detections}
\tablewidth{0pt}
\footnotesize
\tablehead{
\colhead{HzRG} &
\colhead{$\log(L_{\rm H}^{\rm tot}/L_\odot)$} &
\colhead{$\log(L_{\rm H}^{\rm stel}/L_\odot)$} &
\colhead{$f_{\rm stel}$} &
\colhead{$\log(M^{\rm stel}/M_\odot)$} &
\colhead{$\log(L_{5\um}/L_\odot)$} }
\startdata
6C~0032+412 & 12.13 & $ 11.47^{+0.16}_{-0.12}$ & 0.22 & $ 11.28^{+0.16}_{-0.12} $ & $ 11.97^{+0.56}_{-0.24} $ \\ 
6C*0058+495 & 11.16 & $ 11.16^{+0.13}_{-0.10}$ & 0.99 & $ 11.26^{+0.13}_{-0.10} $ & $ 11.40^{+0.16}_{-0.11} $ \\ 
MRC~0350-279 & 11.29 & $ 10.89^{+0.26}_{-0.17}$ & 0.40 & $ 10.85^{+0.26}_{-0.17} $ & $ 10.91^{+0.50}_{-0.23} $ \\ 
MRC~0406-244 & 11.50 & $ 11.49^{+0.13}_{-0.13}$ & 0.98 & $ 11.38^{+0.13}_{-0.13} $ & $ 11.75^{+0.11}_{-0.10} $ \\ 
4C60.07 & 11.67 & $ 11.65^{+0.13}_{-0.10}$ & 0.95 & $ 11.44^{+0.13}_{-0.10} $ & $ 12.50^{+0.35}_{-0.35} $ \\ 
PKS~0529-549 & 11.67 & $ 11.60^{+0.11}_{-0.10}$ & 0.86 & $ 11.46^{+0.11}_{-0.10} $ & $ 11.60^{+0.14}_{-0.10} $ \\ 
4C41.17 & 11.60 & $ 11.60^{+0.08}_{-0.08}$ & 0.99 & $ 11.39^{+0.08}_{-0.08} $ & $ 12.06^{+0.36}_{-0.36} $ \\ 
B2~0902+34 & 11.52 & $ 10.97^{+0.14}_{-0.11}$ & 0.28 & $ 10.81^{+0.14}_{-0.11} $ & $ 11.80^{+0.17}_{-0.12} $ \\ 
USS~0943-242 & 11.40 & $ 11.40^{+0.15}_{-0.07}$ & 0.99 & $ 11.22^{+0.15}_{-0.07} $ & $ 11.67^{+0.35}_{-0.20} $ \\ 
PKS~1138-262 & 12.59 & $< 12.18$  & $ < 0.39 $ & $ <12.11 $ & $ 12.39^{+0.35}_{-0.20} $\\ 
LBDS~53w002 & 11.55 & $ 11.38^{+0.12}_{-0.10}$ & 0.68 & $ 11.27^{+0.12}_{-0.10} $ & $ 11.55^{+0.22}_{-0.14} $ \\ 
3C356.0 & 11.28 & $ 11.28^{+0.12}_{-0.10}$ & 1.00 & $ 11.39^{+0.12}_{-0.10} $ & $ 11.39^{+0.08}_{-0.07} $ \\ 
7C~1751+6809 & 11.14 & $ 11.12^{+0.13}_{-0.13}$ & 0.95 & $ 11.14^{+0.13}_{-0.13} $ & $ 10.39^{+0.30}_{-0.18} $ \\ 
7C~1756+6520 & 11.12 & $ 11.12^{+0.12}_{-0.10}$ & 1.00 & $ 11.15^{+0.12}_{-0.10} $ & $ 10.80^{+0.27}_{-0.17} $ \\ 
3C368.0 & 11.32 & $ 11.32^{+0.11}_{-0.09}$ & 0.99 & $ 11.43^{+0.11}_{-0.09} $ & $ 11.28^{+0.10}_{-0.08} $ \\ 
7C~1805+6332 & 11.22 & $ 11.11^{+0.15}_{-0.11}$ & 0.78 & $ 11.07^{+0.15}_{-0.11} $ & $ 11.19^{+0.17}_{-0.13} $ \\ 
4C40.36 & 11.38 & $ 11.38^{+0.13}_{-0.10}$ & 1.00 & $ 11.29^{+0.13}_{-0.10} $ & $ 11.21^{+0.23}_{-0.15} $ \\ 
TX~J1908+7220 & 12.63 & $< 12.44$  & $ < 0.64 $ & $ <12.27 $ & $ 12.60^{+0.13}_{-0.11} $\\ 
4C23.56 & 11.91 & $ 11.71^{+0.12}_{-0.09}$ & 0.63 & $ 11.59^{+0.12}_{-0.09} $ & $ 12.49^{+0.10}_{-0.09} $ \\ 
B3~J2330+3927 & 12.30 & $< 12.11$  & $ < 0.64 $ & $ <11.94 $ & $ 12.58^{+0.26}_{-0.16} $\\ 
3C470 & 11.32 & $ 11.30^{+0.21}_{-0.15}$ & 0.95 & $ 11.30^{+0.21}_{-0.15} $ & $ 11.80^{+0.11}_{-0.09} $ \\ 
\enddata
\label{table.sedfitting}
\end{deluxetable}

\clearpage
% TABLE 5
\begin{deluxetable}{cccc}
\tablecaption{Results of SED fitting for HzRGs with no MIPS detections}
\tablewidth{0pt}
\footnotesize
\tablehead{
\colhead{HzRG} &
\colhead{$\log(L_{\rm H}^{\rm tot}/L_\odot)$} &
\colhead{$\log(M_{\rm max}^{\rm stel}/M_\odot)$} &
\colhead{$\log(L_{5\um}/L_\odot)$}}
\startdata
MRC~0037-258 & $<11.56$ & $<11.70$ & - \\ 
MRC~0114-211 & $<11.37$ & $<11.40$ & - \\ 
TN~J0121+1320 & 11.21 & 11.00 & - \\ 
6C*0132+330 & 11.11 & 11.10 & - \\ 
6C~0140+326 & $<11.78$ & $<11.40$ & - \\ 
MRC~0152-209 & $<11.72$ & $<11.70$ & - \\ 
MRC~0156-252 & $<12.23$ & $<12.20$ & 12.21 \\ 
TN~J0205+2242 & $<10.96$ & $<10.80$ & - \\ 
MRC~0211-256 & $<11.54$ & $<11.60$ & 11.23 \\ 
3C65 & 11.44 & 11.50 & - \\ 
MRC~0251-273 & 11.14 & 11.00 & - \\ 
MRC~0316-257 & 11.39 & 11.20 & - \\ 
MRC~0324-228 & 11.21 & 11.20 & - \\ 
WN~J0617+5012 & 10.74 & 10.60 & - \\ 
WN~J0747+3654 & 11.34 & 11.20 & - \\ 
6CE0820+3642 & 11.54 & 11.50 & - \\ 
USS~0828+193 & $<11.72$ & $<11.60$ & 12.34 \\ 
5C7.269 & 11.48 & 11.40 & - \\ 
6CE0901+3551 & 11.32 & 11.30 & - \\ 
6CE0905+3955 & 11.40 & 11.40 & - \\ 
TN~J0924-2201 & 11.48 & 11.10 & $ <11.79 $ \\ 
6C~0930+389 & 11.40 & 11.30 & - \\ 
3C239 & 11.60 & 11.60 & - \\ 
MRC~1017-220 & $<11.74$ & $<11.70$ & - \\ 
MG~1019+0534 & $<11.34$ & $<11.20$ & 11.50 \\ 
WN~J1115+5016 & $<10.86$ & $<10.70$ & - \\ 
3C257 & $<11.83$ & $<11.70$ & 11.96 \\ 
WN~J1123+3141 & $<11.81$ & $<11.60$ & - \\ 
%PKS~1138-262 & $<12.30$ & $<12.20$ & 12.41 \\ 
3C266 & 11.22 & 11.30 & - \\ 
6C~1232+39 & $<11.66$ & $<11.50$ & - \\ 
USS~1243+036 & $<11.47$ & $<11.30$ & - \\ 
TN~J1338-1942 & $<11.26$ & $<11.00$ & - \\ 
4C24.28 & $<11.20$ & $<11.00$ & - \\ 
3C294.0 & $<11.50$ & $<11.50$ & - \\
USS~1410-001 & $<11.53$ & $<11.40$ & 11.86 \\ 
8C~1435+635 & $<11.45$ & $<11.10$ & $ <11.63 $ \\ 
USS~1558-003 & 11.85 & 11.70 & 11.83 \\ 
USS~1707+105 & 11.28 & 11.20 & - \\ 
%LBDS~53w069 & 11.09 & 11.10 & $ <10.74 $ \\ 
LBDS~53w091 & 11.20 & 11.20 & $ <10.88 $ \\ 
WN~J1911+6342 & $<11.13$ & $<10.90$ & $ <11.33 $ \\ 
TN~J2007-1316 & $<11.79$ & $<11.60$ & - \\ 
MRC~2025-218 & 11.80 & 11.60 & 11.53 \\ 
MRC~2048-272 & $<11.55$ & $<11.50$ & - \\ 
MRC~2104-242 & 11.34 & 11.20 & 11.30 \\ 
MG~2144+1928 & $<11.48$ & $<11.30$ & - \\ 
USS~2202+128 & $<11.85$ & $<11.70$ & 11.77 \\ 
MRC~2224-273 & $<11.34$ & $<11.30$ & - \\ 
4C28.58 & $<11.56$ & $<11.40$ & 11.68 
\enddata
\label{table.sedfitting2}
\end{deluxetable}
%\addtocounter{table}{-1}

% TABLE 5
%\begin{deluxetable}{cccc}
%\tablecaption{(continued) Results of SED fitting for HzRGs with no MIPS detections}
%\tablewidth{0pt}
%\footnotesize
%\tablehead{
%\colhead{HzRG} &
%\colhead{$\log(L_{\rm H}^{\rm tot}/L_\odot)$} &
%\colhead{$\log(M_{\rm max}^{\rm stel}/M_\odot)$} &
%\colhead{$\log(L_{5\um}/L_\odot)$}}
%\startdata
%\input{tab5b}
%\enddata
%\end{deluxetable}

\clearpage

% FIGURE 1
\clearpage

% FIGURE 2
\clearpage

\clearpage
\clearpage
\clearpage
\begin{center}
  \begin{minipage}{0.33\linewidth}
    \includegraphics[width=5.41cm,angle=270]{f5g.ps}
  \end{minipage}\hspace*{15mm}
  \begin{minipage}{0.33\linewidth}
    \includegraphics[width=5.41cm,angle=270]{f5h.ps}
  \end{minipage}
  \begin{minipage}{0.33\linewidth}
    \includegraphics[width=5.41cm,angle=270]{f5i.ps}
  \end{minipage}\hspace*{15mm}
  \begin{minipage}{0.33\linewidth}
    \includegraphics[width=5.41cm,angle=270]{f5j.ps}
  \end{minipage}\\
 \begin{minipage}{0.33\linewidth}
    \includegraphics[width=5.41cm,angle=270]{f5k.ps}
  \end{minipage}\hspace*{15mm}
  \begin{minipage}{0.33\linewidth}
    \includegraphics[width=5.41cm,angle=270]{f5l.ps}
  \end{minipage}\\
   \begin{minipage}{0.33\linewidth}
    \includegraphics[width=5.41cm,angle=270]{f5m.ps}
  \end{minipage}\hspace*{15mm}
  \begin{minipage}{0.33\linewidth}
    \includegraphics[width=5.41cm,angle=270]{f5n.ps}
  \end{minipage}\\
  \begin{minipage}[r]{0.33\linewidth}
    \includegraphics[width=5.41cm,angle=270]{f5o.ps}
  \end{minipage}\hspace*{15mm}
  \begin{minipage}{0.33\linewidth}
    \includegraphics[width=5.41cm,angle=270]{f5p.ps}
  \end{minipage}\\
  \begin{minipage}{0.33\linewidth}
    \includegraphics[width=5.41cm,angle=270]{f5q.ps}
  \end{minipage}\hspace*{15mm}
  \begin{minipage}{0.33\linewidth}
    \includegraphics[width=5.41cm,angle=270]{f5r.ps}
  \end{minipage}\\
  \begin{minipage}{0.33\linewidth}
    \includegraphics[width=5.41cm,angle=270]{f5s.ps}
  \end{minipage}\hspace*{15mm}
  \begin{minipage}{0.33\linewidth}
    \includegraphics[width=5.41cm,angle=270]{f5t.ps}
  \end{minipage}\\
  \begin{minipage}{0.33\linewidth}
    \includegraphics[width=5.41cm,angle=270]{f5u.ps}
  \end{minipage}\hspace*{15mm}
\end{center}
\clearpage
\clearpage
\begin{center}
  \begin{minipage}{0.33\linewidth}
    \includegraphics[width=5.41cm,angle=270]{f6g.ps}
  \end{minipage}\hspace*{15mm}
  \begin{minipage}{0.33\linewidth}
    \includegraphics[width=5.41cm,angle=270]{f6h.ps}
  \end{minipage}\\
  \begin{minipage}{0.33\linewidth}
    \includegraphics[width=5.41cm,angle=270]{f6i.ps}
  \end{minipage}\hspace*{15mm}
  \begin{minipage}{0.33\linewidth}
    \includegraphics[width=5.41cm,angle=270]{f6j.ps}
  \end{minipage}\\
  \begin{minipage}{0.33\linewidth}
    \includegraphics[width=5.41cm,angle=270]{f6k.ps}
  \end{minipage}\hspace*{15mm}
  \begin{minipage}{0.33\linewidth}
    \includegraphics[width=5.41cm,angle=270]{f6l.ps}
  \end{minipage}\\
  \begin{minipage}{0.33\linewidth}
    \includegraphics[width=5.41cm,angle=270]{f6m.ps}
  \end{minipage}\hspace*{15mm}
  \begin{minipage}{0.33\linewidth}
    \includegraphics[width=5.41cm,angle=270]{f6n.ps}
  \end{minipage}\\
  \begin{minipage}[r]{0.33\linewidth}
    \includegraphics[width=5.41cm,angle=270]{f6o.ps}
  \end{minipage}\hspace*{15mm}
  \begin{minipage}{0.33\linewidth}
    \includegraphics[width=5.41cm,angle=270]{f6p.ps}
  \end{minipage}\\
  \begin{minipage}{0.33\linewidth}
    \includegraphics[width=5.41cm,angle=270]{f6q.ps}
  \end{minipage}\hspace*{15mm}
  \begin{minipage}{0.33\linewidth}
    \includegraphics[width=5.41cm,angle=270]{f6r.ps}
  \end{minipage}\\
  \begin{minipage}{0.33\linewidth}
    \includegraphics[width=5.41cm,angle=270]{f6s.ps}
  \end{minipage}\hspace*{15mm}
  \begin{minipage}{0.33\linewidth}
    \includegraphics[width=5.41cm,angle=270]{f6t.ps}
  \end{minipage}\\
  \begin{minipage}{0.33\linewidth}
    \includegraphics[width=5.41cm,angle=270]{f6u.ps}
  \end{minipage}\hspace*{15mm}
  \begin{minipage}{0.33\linewidth}
    \includegraphics[width=5.41cm,angle=270]{f6v.ps}
  \end{minipage}\\
  \begin{minipage}{0.33\linewidth}
    \includegraphics[width=5.41cm,angle=270]{f6w.ps}
  \end{minipage}\hspace*{15mm}
  \begin{minipage}{0.33\linewidth}
    \includegraphics[width=5.41cm,angle=270]{f6x.ps}
  \end{minipage}\\
  \begin{minipage}{0.33\linewidth}
    \includegraphics[width=5.41cm,angle=270]{f6y.ps}
  \end{minipage}\hspace*{15mm}
  \begin{minipage}{0.33\linewidth}
    \includegraphics[width=5.41cm,angle=270]{f6z.ps}
  \end{minipage}\\
  \begin{minipage}{0.33\linewidth}
    \includegraphics[width=5.41cm,angle=270]{f6aa.ps}
  \end{minipage}\hspace*{15mm}
  \begin{minipage}{0.33\linewidth}
    \includegraphics[width=5.41cm,angle=270]{f6ab.ps}
  \end{minipage}\\
  \begin{minipage}{0.33\linewidth}
    \includegraphics[width=5.41cm,angle=270]{f6ac.ps}
  \end{minipage}\hspace*{15mm}
  \begin{minipage}[r]{0.33\linewidth}
    \includegraphics[width=5.41cm,angle=270]{f6ad.ps}
  \end{minipage}\\
  \begin{minipage}{0.33\linewidth}
    \includegraphics[width=5.41cm,angle=270]{f6ae.ps}
  \end{minipage}\hspace*{15mm}
  \begin{minipage}{0.33\linewidth}
    \includegraphics[width=5.41cm,angle=270]{f6af.ps}
  \end{minipage}\\
  \begin{minipage}{0.33\linewidth}
    \includegraphics[width=5.41cm,angle=270]{f6ag.ps}
  \end{minipage}\hspace*{15mm}
  \begin{minipage}{0.33\linewidth}
    \includegraphics[width=5.41cm,angle=270]{f6ah.ps}
  \end{minipage}\\
  \begin{minipage}{0.33\linewidth}
    \includegraphics[width=5.41cm,angle=270]{f6ai.ps}
  \end{minipage}\hspace*{15mm}
  \begin{minipage}{0.33\linewidth}
    \includegraphics[width=5.41cm,angle=270]{f6aj.ps}
  \end{minipage}\\
  \begin{minipage}{0.33\linewidth}
    \includegraphics[width=5.41cm,angle=270]{f6ak.ps}
  \end{minipage}\hspace*{15mm}
  \begin{minipage}{0.33\linewidth}
    \includegraphics[width=5.41cm,angle=270]{f6al.ps}
  \end{minipage}\\
  \begin{minipage}{0.33\linewidth}
    \includegraphics[width=5.41cm,angle=270]{f6am.ps}
  \end{minipage}\hspace*{15mm}
  \begin{minipage}{0.33\linewidth}
    \includegraphics[width=5.41cm,angle=270]{f6an.ps}
  \end{minipage}\\
  \begin{minipage}{0.33\linewidth}
    \includegraphics[width=5.41cm,angle=270]{f6ao.ps}
  \end{minipage}\hspace*{15mm}
  \begin{minipage}{0.33\linewidth}
    \includegraphics[width=5.41cm,angle=270]{f6ap.ps}
  \end{minipage}\\
  \begin{minipage}{0.33\linewidth}
    \includegraphics[width=5.41cm,angle=270]{f6aq.ps}
  \end{minipage}\hspace*{15mm}
  \begin{minipage}{0.33\linewidth}
    \includegraphics[width=5.41cm,angle=270]{f6ar.ps}
  \end{minipage}\\
  \begin{minipage}[r]{0.33\linewidth}
    \includegraphics[width=5.41cm,angle=270]{f6as.ps}
  \end{minipage}\hspace*{15mm}
  \begin{minipage}{0.33\linewidth}
    \includegraphics[width=5.41cm,angle=270]{f6at.ps}
  \end{minipage}\\
  \begin{minipage}{0.33\linewidth}
    \includegraphics[width=5.41cm,angle=270]{f6au.ps}
  \end{minipage}\hspace*{15mm}
  \begin{minipage}{0.33\linewidth}
    \includegraphics[width=5.41cm,angle=270]{f6av.ps}
  \end{minipage}\\
\end{center}

\clearpage

\clearpage

\clearpage

\clearpage

\clearpage

\clearpage

\clearpage

\clearpage

\end{document}